\documentclass[11pt]{aastex}
\usepackage{emulateapj5,apjfonts}
\usepackage{onecolfloat}
\usepackage{epsf}

\slugcomment{{\sc To appear in ApJ:} 20 March 2001}

\newcommand{\ml}{{$M/L$} }
\newcommand{\mls}{{$M/L$s} }
\newcommand{\mln}{$M/L$}
\newcommand{\mlsn}{$M/L$s}
\newcommand{\hi}{H{\sc i} }

\shorttitle{Stellar M/Ls and the Tully-Fisher relation}
\shortauthors{Bell \& de Jong}

\begin{document}

\def\head{
\title{Stellar Mass-to-light Ratios and the Tully-Fisher Relation}
\author{Eric F.~Bell and Roelof S.~de Jong\altaffilmark{1}}
\affil{Steward Observatory, University of Arizona, 933 N. Cherry Ave., 
	Tucson, AZ 85721, USA}
\email{ebell,rdejong@as.arizona.edu}
\altaffiltext{1}{Hubble fellow}

\begin{abstract}
We have used a suite of simplified spectrophotometric spiral galaxy 
evolution models to argue that there are substantial variations in 
stellar mass-to-light ratios (\mlsn) within and among galaxies, amounting 
to factors of between 3 and 7 
in the optical, and factors of 2 in the near-infrared.
Our models show a strong correlation between stellar \ml and the optical
colors of the integrated stellar populations.
Under the assumption of a universal spiral galaxy IMF, 
relative trends in model stellar \ml with color are robust to 
uncertainties in stellar population and galaxy evolution modeling, including
the effects of modest bursts of star formation.
Errors in the dust reddening estimates do not strongly affect the final 
derived stellar masses of a stellar population.  
We examine the observed maximum disk stellar \mls of a sample 
of spiral galaxies with accurate rotation curves and optical and
near-infrared luminosity profiles. From these observed maximum disk \mls we
conclude that a Salpeter Initial Mass Function (IMF) has too many low-mass
stars per unit luminosity, but that an IMF similar to the Salpeter IMF
at the high-mass end with less low-mass stars (giving stellar \mls 30\% 
lower than the Salpeter value) is consistent
with the maximum disk constraints.  Trends in observed 
maximum disk stellar \mls
with color provide a good match to the predicted model relation, suggesting
that the spiral galaxy stellar IMF is universal and that a fraction of
(particularly high surface brightness) 
spiral galaxies may be close to maximum disk.
We apply the model trends in stellar \ml with color to the
Tully-Fisher (TF) relation.
We find that the stellar mass TF relation is relatively steep and has modest
scatter, and 
is independent of the passband and color used to derive the stellar masses, 
again lending support for a universal IMF. 
The difference in slope between the optical (especially blue) and 
near-infrared TF relations 
is due to the combined effects of dust attenuation
and stellar \ml variations with galaxy mass.  
Assuming the HST Key Project distance to the Ursa Major 
Cluster and neglecting the (uncertain) molecular gas fraction,
we find that the baryonic TF relation 
takes the form $M_{\rm baryon} \propto V^{3.5}$ (with 
random and systematic 1$\sigma$ slope errors of $\sim$ 0.2 each)
when using a bisector fit and rotation 
velocities derived from the flat part of the rotation curve.
Since we have normalized the stellar \mls to be as high as can possibly
be allowed by maximum disk constraints, the slope of the baryonic TF relation 
will be somewhat shallower than 3.5
if all disks are substantially sub-maximal.
\end{abstract}

\keywords{
galaxies : spiral --- 
galaxies : stellar content --- 
galaxies : evolution --- 
galaxies : fundamental parameters
}

}%%%end head

\twocolumn[\head]

%----------------------------------------------------------------------
\section{INTRODUCTION}

The stellar mass-to-light ratio (\mln) is an important
parameter in astrophysics as it allows translation between
photometry and dynamics.  The stellar \ml has a direct bearing
on two hotly debated areas in spiral galaxy research: the appropriate
stellar \mls to be used for spiral galaxy rotation curve decompositions, and 
the passband-dependent slope of the galaxy magnitude-rotation velocity
relation \citep[TF relation hereafter]{tully77}.
In this paper, we address the stellar \mls of spiral galaxies,
briefly explore the implications of our results for rotation curve
decompositions and investigate in more depth the slope of the 
TF relation.

There is presently much interest in decomposing spiral galaxy rotation curves
into contributions from the gaseous, stellar and dark matter contents
\citep[e.g.][]{verheijen97,deblok98}.
The primary motivation for this interest is that, in principle, the 
structure of dark matter halos can be determined from spiral galaxy rotation
curves {\it if} the contribution from gas and stars can be properly understood.
In turn, the structure of dark matter halos is a strong constraint
on dark matter halo formation models \citep[e.g.][]{moore98,navarro00a}.  
The main challenge in determining
the dark matter contribution to a given rotation curve is our ignorance
regarding plausible values of the stellar \mln: the gas contribution is 
typically well-understood and relatively small \citep{verheijen97,swaters00}.  
The situation is degenerate enough that many
rotation curves can be equally well-fit by models in which 
the central parts of the rotation curve are dominated entirely by 
stellar mass or by dark matter 
\citep[e.g.][]{vanalbada85,swaters99}.  In order to resolve this 
degeneracy, some independent constraints on stellar \mlsn,
and their variations with radius and galaxy properties, are
required.

The implications of the stellar \ml
for the TF relation are no less important.  The TF 
relation relates the integrated luminosity in a given passband
to the global dynamics of the galaxy and its dark matter halo.
The dust-corrected TF relation has a slope
which steepens towards redder passbands \citep[going between 
$L \propto V^3$ or shallower in the optical to 
$L \propto V^4$ in the near-infrared; ][]{verheijen97,tully98}
indicating that there is a trend in color and stellar
\ml with galaxy mass.  This change in slope with passband can 
considerably weaken the power of the TF relation as a 
test of galaxy formation and evolution models 
\citep[such as those by][]{cole00,navarro00b,vdbosch00}:
it is possible to reproduce the TF relation in one passband
easily without reproducing the TF relation in other passbands
\citep[for a multi-waveband comparison of models with 
the TF relation see e.g.][]
{heavens99}.

One way around this confusion is to explore the total 
baryonic mass TF relation.
An estimate of the baryonic TF relation can be obtained by
adding the gas mass to a crude estimate of stellar mass implied by the
luminosity (usually assuming a constant \mln).
This has been attempted the most thoroughly by \citet{mcgaugh00}
using a constant \ml in $B$, $I$, $H$, and $K$ bands, 
although e.g.\ \citet{milgrom88} and \citet{matthews98} discussed
aspects of this problem.  Because the stellar
\ml is likely to vary along the TF relation in all passbands, their composite 
baryonic TF relation will have a larger scatter and different 
slope than the true baryonic TF relation.
A deeper and firmer understanding
of the baryonic TF relation is only possible once variations 
in stellar \ml along the TF relation are understood and 
incorporated in the analysis.

In this paper, we use simplified spiral galaxy evolution models
similar to the ones presented by \citet{papiii}
to investigate plausible trends in stellar \ml with galaxy properties,
assuming a universal IMF.
We discuss these models briefly in \S \ref{sec:mod}.  In 
\S \ref{sec:ml} we investigate trends in spiral galaxy stellar
\ml for a number of plausible models, finding that there are systematic
variations in stellar \ml as a function of many galaxy parameters, and
that stellar \mls correlate most tightly with galaxy color.  
In \S \ref{sec:unc} we investigate the physical basis of
the color--\ml relation and
we discuss uncertainties in the stellar \mlsn, including the effects of 
using different stellar population models, different IMFs, 
different galaxy evolution prescriptions, and dust.
In \S \ref{sec:rotn} we discuss the implications
of these variations in stellar \ml for rotation curve decompositions, 
and put the stellar \mls onto an observationally-determined 
maximum-disk scale.  
In \S \ref{sec:tf} we then discuss
at length the implications of these variations in stellar
\ml for the stellar mass and baryonic TF relation.  
Finally, in \S \ref{sec:conc}, we present our conclusions.
Readers not interested in the details of the models and a detailed
analysis of the uncertainties in model stellar \mls can skip 
\S\S \ref{sec:mod} and \ref{sec:unc}.
Note that we state all stellar \mls in solar units.
We adopt the HST Key Project distance scale in 
this paper, corresponding to $H_0 = 71$ km\,s$^{-1}$\,Mpc$^{-1}$
\citep{sakai00}.

%----------------------------------------------------------------------
\section{THE GALAXY EVOLUTION MODELS} \label{sec:mod}

To construct the model \mls for spiral
galaxies, we use models similar to those presented by
\citet{papiii}.  They presented a suite of 
simple spectrophotometric disk evolution models designed to reproduce
many of the trends between the radially-resolved colors of
spiral galaxies and their structural parameters, as observed
by \citet{papii}.  These models were not designed to 
address the evolution of bulges or dwarf Spheroidal 
galaxies: the star formation laws used in these models 
(parameterized using surface density) 
are valid only for disk-dominated galaxies.
These models describe the evolution 
of a gaseous disk, according to a prescribed
star formation law and chemical evolution prescription (assuming
the instantaneous recycling approximation; IRA).
Relaxing the IRA would have two effects: it would allow non-solar
abundance ratios to develop, and it would slightly modify
the time evolution of the metallicity of galaxies.
Most stellar population models are incapable of dealing 
adequately with non-solar abundance ratios: however, it looks likely
that the effects of non-solar abundance ratios
on integrated colors are modest \citep[as they mimic the effects
of modest changes in metallicity; e.g.][]{alpha}.
Furthermore, the time evolution of spiral galaxy metallicity 
(which is dominated, by mass, by the Type {\sc ii} supernova 
product oxygen) is described fairly accurately by the IRA
except at late stages of galactic evolution near gas exhaustion
\citep[e.g.][]{tinsley80,pagel98,portinari99,prantzos00}.
Thus, our use of the IRA is a reasonable approximation, 
bearing in mind the modest effects caused by adopting it, 
and the considerable stellar population and galaxy
evolution modeling uncertainties.

To construct radially-resolved 
stellar population colors, the stellar populations
synthesis (SPS) models of \citet{bruzual00}, as described in \citet{liu00} are
used, adopting a \citet{salpeter55} IMF, which we modify 
by globally scaling down
its stellar \ml by a factor of 0.7 \citep[cf.][]{baryon}.  We adopt lower
and upper mass limits of 0.1M$_{\sun}$ and 125M$_{\sun}$ respectively.
In \citet{papiii} we adopted a pure
\citet{salpeter55} IMF: in this paper, we have been forced to adopt an IMF
with lower \mls to agree with observational maximum disk 
\ml constraints (see \S \ref{sec:rotn}).
This global reduction in stellar \ml is essentially the same 
as adopting an IMF with with fewer low-mass stars, as the low mass
stars contribute only to the mass, but not the luminosity or color, of
the stellar population.  It is interesting to note that 
there is increasing empirical evidence for a universal
IMF with a Salpeter slope for stars more massive 
than the Sun, and a shallower slope for stars less massive than 
the Sun \citep{kroupa00}.  
This IMF has stellar \mls comparable to or slightly
lower than the maximum disk-scaled IMF we adopt in this paper.
It is important to note that neither the slope nor the scatter
of the stellar \mlsn, nor the trends in color
with galaxy properties are affected by 
our adoption of a scaled-down Salpeter IMF: the only effect on 
the following analysis is to modify the overall normalization of 
the stellar \mlsn.  

For our models, we follow the evolution of 
an exponential gaseous disk using either a \citet{schmidt59}
local gas density-dependent star formation law or a 
gas density- and dynamical time-dependent star formation
law \citep{kennicutt98}.  Model galaxies with a wide range 
of masses and central surface densities are generated, as
we do not attempt to {\it a priori} predict 
the mass and central surface density distributions of spiral galaxies.
To avoid comparing the observed galaxies to model galaxies without
any observed analogue from \citet{papii}, 
we select model galaxies to have a similar
range in $K$ band absolute magnitudes and 
central surface brightnesses as their observed galaxies (including 
an observed modest absolute magnitude--central 
surface brightness correlation).
These models are tuned to reproduce observed
trends in color-based local age and metallicity as a function 
of local $K$ band surface brightness, in conjunction with 
the observed correlation between gas fraction and 
$K$ band central surface brightness \citep{papii,papiii}.

We present a total of six models in this paper.  i) We first use a closed
box model, with no gas infall or outflow, a galaxy age
of 12 Gyr and a Schmidt star formation law.  The main disadvantages
of this model is the lack of a strong metallicity-magnitude correlation and 
weaker age--magnitude correlation, and the underprediction of the age 
gradients.
We then allow ii) gas infall (whose timescale depends on 
galaxy mass and radius) {\it or} iii) metal-enriched outflow, both of which 
alleviate the above shortcomings of the closed-box model.  iv) We 
then adopt a dynamical time-dependent star formation law (without
infall or outflow), which 
we find produces a `backwards' metallicity--magnitude correlation, and 
is therefore unacceptable, in isolation.  v) We then explore
the use of a mass-dependent galaxy formation epoch without infall or
outflow, which imprints
metallicity--magnitude and age--magnitude correlations.  A mass-dependent
formation epoch is a common feature of many cosmologically-motivated
galaxy formation models \citep[e.g.][]{somerville99,cole00}. vi) Finally,
we explore a `burst' model with a mass-dependent galaxy formation 
epoch and no infall or outflow, 
where the star formation rate is varied on 0.5 Gyr timescales
with a log-normal distribution with a factor-of-two width.
None of these models perfectly describe the trends in spiral galaxy 
colors with galaxy parameters observed in \citet{papii}; however, the models
taken as a suite encompass the range of behaviors seen in the
observed galaxy sample.  We adopt the mass-dependent formation epoch 
model with bursts, with a scaled-down Salpeter IMF, as the default model.
This model reproduces the trends in local spiral galaxy 
age and metallicity with 
local $K$ band surface brightness with acceptable scatter, while
simultaneously reproducing the age--magnitude and metallicity--magnitude 
correlations with acceptable scatter.
However, as we later demonstrate (see e.g.\ \S \ref{subsec:galevo}
and Fig.\ \ref{fig:amod}), the choice of model does not significantly
affect any of our conclusions.
For more model details, see \citet{papiii}.

%----------------------------------------------------------------------
\section{CONSTRUCTING MODEL MASS-TO-LIGHT RATIOS} \label{sec:ml}

We use the spiral galaxy evolution models (which reproduce
the trends in spiral galaxy color with structural parameters) to
construct stellar \mls for integrated stellar populations.
These are converted into solar units assuming
solar absolute magnitudes of 5.47, 4.82, 4.46, 4.14, and 3.33 in
Johnson $B$ and $V$, Kron-Cousins $R$ and $I$, and Johnson $K$ passbands 
respectively \citep{cox00,bessel79}.
We also adopt Johnson $J$ and $H$ band solar absolute magnitudes of
3.70 and 3.37 respectively 
from \citet{worthey94} as \citet{cox00} does not present $J$ and 
$H$ band magnitudes of the Sun: \citet{worthey94} magnitudes in 
other passbands are comparable to those presented by \citet{cox00}.
Instead of using the full gas mass loss histories from the
SPS models, we used the IRA to construct the stellar masses.
This may lead to errors of $\lesssim 5$ per
cent in stellar \ml (compared to the exact value).
Bearing in mind the 
size of variations in \ml the model predicts (greater than a factor of 
two), and the other considerable uncertainties affecting the stellar 
\mlsn, such as the stellar initial mass function (IMF) and dust, our
use of the IRA is more than acceptable.

%figure:ml
\begin{figure}[tbh]
\epsfxsize=\linewidth
\epsfbox[72 190 527 650]{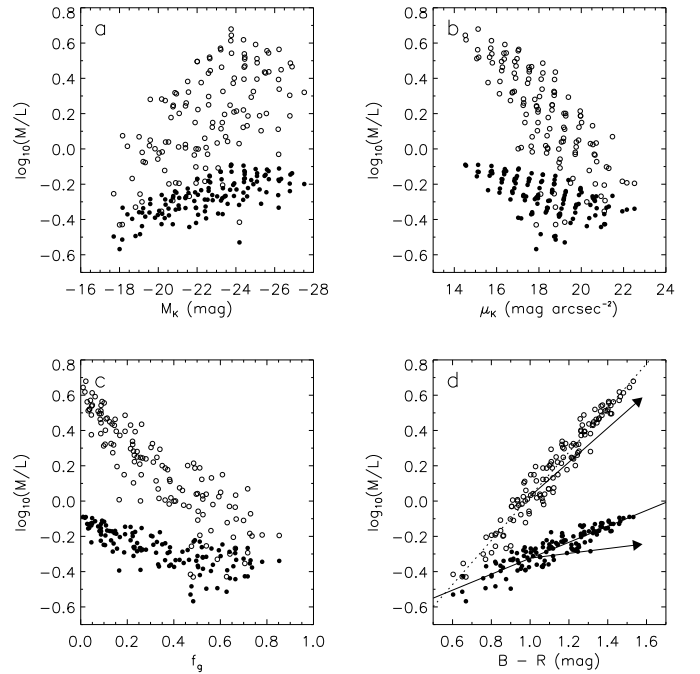}
\caption{\label{fig:ml} 
Trends in model stellar \protect\mls with
galaxy parameters for the formation epoch model with
bursts.  We show the trends in model stellar \ml in the 
$B$ band (open circles) and $K$ band (filled circles) as a
function of $K$ band absolute magnitude (a), $K$ band 
central surface brightness (b), gas fraction (c), 
and $B - R$ galaxy color (d).  In panel d, we also 
show the fit to the variation of model stellar \ml 
with $B$--$R$ color for this model in $B$ 
(dotted line) and $K$ (solid line) 
and dust extinction vectors in $B$ and $K$ band (arrows)
following \protect\citet{tully98}.  The dust extinction
vectors represent the correction to face-on suffered
by a Milky Way-type galaxy viewed at an inclination 
of 80 degrees.\vspace{-0.2cm}
}
\end{figure}

We show an example of the stellar \mls of our model galaxies for the
mass-dependent formation epoch with bursts 
model in Fig.\,\ref{fig:ml}.  We show
this particular model for two reasons.  Firstly, this model
provides the best match to the overall observed galaxy properties.
Secondly, and more importantly, this model shows the most scatter of any of 
our models, but has quantitatively the same overall behavior as all of 
our models (see e.g.\ \S \ref{subsec:galevo}
and Fig.\ \ref{fig:amod}).  We show the trends in stellar \ml in the 
$B$ band (open circles) and $K$ band (filled circles) as a
function of $K$ band absolute magnitude (a), $K$ band 
central surface brightness (b), gas fraction (c), 
and $B$--$R$ galaxy color (d).  Results for other models are presented
in Appendix \ref{sec:app}.

One obvious conclusion is that there are significant trends in model
stellar \ml with all four depicted galaxy parameters {\it in all passbands, 
even in the $K$ band}.  The trends amount to factors of $\sim 7$ in 
$B$, $\sim 3$ in $I$, and $\sim 2$ in $K$ for plausible ranges
of galaxy parameters.  This firmly dispels the notion of a constant
stellar \ml for a spiral galaxy in any passband:  this conclusion
is even true in $K$ band, where there have been claims that the
stellar \ml will be robust to differences in star formation history
\citep[SFH; e.g.][]{djiv,verheijen97}.
Of course, we find that the trends in stellar \ml are minimized
in $K$ band: this suggests that $K$ band observations are important
for any observations in which minimizing scatter in \ml is 
important (e.g.\ for rotation curve studies).  

The scatter in model stellar \ml at a given magnitude is rather large as a
consequence of the modeling assumptions. The SFH of
our model galaxies depends primarily on their local surface density, and
only weakly on their total mass, as is observed \citep{papii}.  As
galaxies come in a range of surface brightnesses at a given magnitude
\citep{dejong00}, our models will have a range of SFHs and consequently \mls
at a given magnitude. There is considerable scatter in stellar
\ml with $K$ band central surface brightness and with gas fraction; however,
this scatter is highly model dependent as there is no scatter in these 
relations for the closed box models, and intermediate scatter for the 
outflow and infall models.  

One important conclusion is that, for all the models investigated for
this paper, the model stellar \mls in all optical and 
near-infrared (near-IR) passbands
correlate strongly, with minimal scatter, with galaxy color 
\citep[see also][]{bottema97}.  This is expected: 
the star formation and chemical enrichment history determine
both the stellar \ml and galaxy color. 
Later, we demonstrate that the slope of the stellar \mln--color
correlation is very robust, and we place a strong constraint on the 
zero point of the correlation.  
%Thus, the stellar 
%\ml of a stellar population can be estimated 
%given an accurate e.g.\ $B - V$ or $B - R$ color
%and a realistic galaxy evolution model, at least in a relative
%sense.  Furthermore, the stellar \ml can be placed on a maximum
%disk scale, or a sub-maximal disk scale of the user's choice (by simply
%scaling down the maximal disk case by an appropriate factor).
This correlation is a powerful tool for understanding 
stellar \mls of spiral galaxies for use in e.g.\ rotation 
curve decompositions or in constructing passband-independent 
TF relations.  We tabulate least-squares fits to the maximum
disk-scaled color--stellar
\ml relations in Table \ref{tab:mlcol} of 
Appendix\,\ref{sec:app} for all models
introduced in \S\ref{sec:mod} and for a broad range in color
combinations.  

Using our models we predict, {\it under the assumption of a universal IMF},
that workers determining the stellar \mls of spiral
galaxies \citep[e.g.][]{bottema93,bottema99,swaters99,weiner00} 
will, with sufficient 
sample size and control of the systematic uncertainties, 
observe trends in stellar \ml which correlate most tightly with galaxy color. 
In \S \ref{sec:rotn}, we demonstrate that there are 
already indications from rotation curve studies that the
correlation between \ml and color has been observed 
\citep[see also][]{ratnam00}.

Another interesting implication of the tight correlation between
stellar \ml and color is that, because color gradients are common in
spiral galaxies, significant gradients in stellar \ml should be
present in most spirals, in the sense that the outer regions of
galaxies will tend to have lower stellar \ml than the inner regions of
galaxies (assuming a universal IMF).  
Obviously this stellar \ml gradient will vary on a
case-by-case basis.  For many galaxies the assumption of a constant
stellar \ml over the disk will not significantly 
affect mass decompositions using
rotation curves, as in the outer regions (where the stellar \ml
is lower) the stars contribute much less to the total mass than the
dark matter \citep[e.g.][]{weiner00}. 
Nevertheless, for accurate rotation curve studies, or
studies based on e.g.\ $B$ band photometry where the stellar \ml
varies strongly as a function of color, the radial variation of
stellar \ml should not be ignored lightly.  A detailed study 
of spiral galaxy rotation curves, using these model stellar 
\mlsn, will be presented in our next paper.

%--------------------------------------------------------------------------
\section{HOW ROBUST ARE THE STELLAR MASS-TO-LIGHT RATIOS?} \label{sec:unc}

In the previous section, we made some strong claims about the stellar
\mls of spiral galaxies.  However, there are a number of uncertainties
which may affect the model stellar \mlsn, 
such as uncertainties in SPS 
and galaxy evolution models, dust, and most importantly, 
the stellar IMF (and possible trends in IMF with galaxy type and structure).  
In the next sections, we discuss
some of these uncertainties and the bearing of these on our results.

%figure:grids
\begin{figure}[tbh]
\epsfxsize=\linewidth
\epsfbox[20 165 562 688]{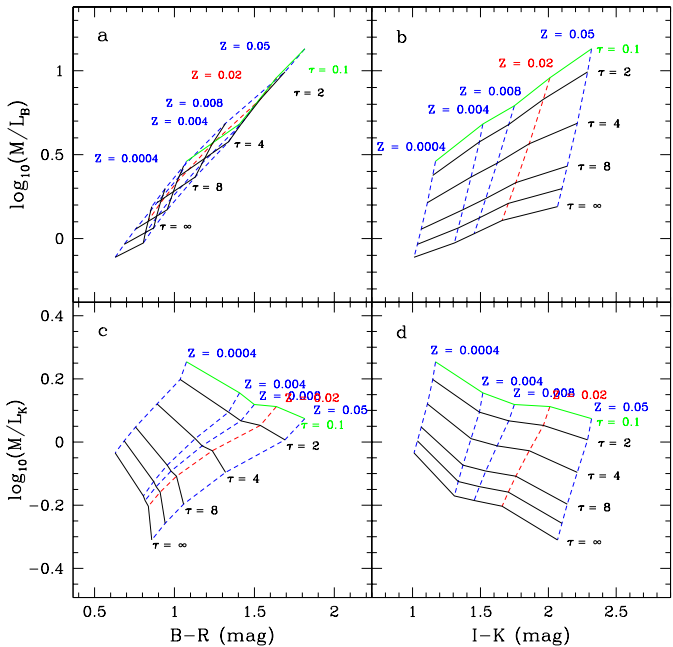}
\caption{\label{fig:grids}
Trends in simple exponential SFH model
stellar \mls with color.  Stellar \mls 
for a Salpeter IMF in $B$ (panels a and b) and
$K$ (panels c and d) of single metallicity exponentially
declining star formation rate models from \citet{bruzual00}
are shown against the model $B - R$ (panels a and c) 
and $I - K$ (panels b and d) broadband colors.
Models of the same $e$-folding time scale
$\tau$ have been connected by solid lines, models of the same
metallicity $Z$ are connected by dashed lines. \vspace{-0.2cm}
}
\end{figure}

\subsection{The origin of the color--\ml correlation}

Before we can assess the uncertainties in the model 
color--\ml relations, we have to understand 
why the correlation between color and stellar \ml exists 
in the first place. To this end, we show in
Fig.\,\ref{fig:grids} color versus stellar \ml for a grid of exponentially
declining star formation rate models.  To construct model colors, we
use SPS models with different
metallicities from \citet{bruzual00}. We use exponentially
declining star formation rates as models with this type of 
SFH can reproduce the optical--near-IR colors of spiral
galaxies quite naturally \citep[e.g.][]{papii}.  Furthermore, 
a slowly declining or constant SFH is inferred for the solar
neighborhood \citep[e.g.][]{rocha00}.  
The exponential decline in star formation rate is
parametrized by $e$-folding timescale $\tau$ and the colors and \mls are
evaluated after a lifetime of 12\,Gyr.  Models with different
$\tau$ but the same metallicity are connected by solid lines, the same
$\tau$s but different metallicities are connected by dashed
lines.

When we consider the model grid for \ml in the $B$ band versus $B - R$
color (Fig.\,\ref{fig:grids}, panel a), we can immediately 
see why the $B$ band stellar \mln-color relation works so well.
There is a tight correlation between $B - R$ color and stellar \ml 
independent of metallicity or SFH.
Similar results are obtained for \mls in other optical 
passbands in combination with optical-optical colors.

The situation is slightly more complex when looking at trends in 
the $K$ band stellar \ml with optical color (Fig.\,\ref{fig:grids}, 
panel c). The age (as parameterized by $\tau$) and metallicity effects
are no longer degenerate. However, realizing that chemical evolution
caused by modest amounts of star formation raise the galaxy metallicity
rapidly to at least 1/10th solar ($Z=0.002$; in a closed box, 
conversion of $\sim$20\%
of the gas mass into stars raises the average
stellar metallicity to over 0.1 solar), 
the range of relevant metallicities becomes narrower, and 
the color--\ml correlation becomes tighter.
Still, we expect a bit more scatter in the relations in the
$K$ band, in particular for very young galaxies with nearly primordial
metallicities \citep[like SBS 1415$+$437: with a metallicity of 0.05 solar it
is one of the lowest metallicity galaxies known; ][]{thuan99}.

We see that the method definitely breaks down when using $I - K$ versus
\ml (Fig.\,\ref{fig:grids}, panels b and d).  This is because we are now
using a color that is mainly a metallicity tracer versus \mln, which is
more sensitive to age effects.  We therefore expect the method to work
best with optical-optical color combinations (which are unfortunately
most affected by dust).  Even though the $K$ band \ml--color relations are
less tight, because of its much smaller dynamic range it is still the
passband prefered for mass estimates, with $I$ band providing an useful
alternative. 

\subsection{Stellar population model uncertainties and IMFs}

In the above analysis we used the SPS models of
\citet{bruzual00} with a scaled-down Salpeter IMF, in conjunction
with our own simple galaxy evolution models, to probe trends in stellar
\ml with galaxy properties.  However, the SPS models
carry with them their own sets of uncertainties, such as the
prescriptions for post-main sequence evolution and the relationship
between stellar properties and the observable colors.  For this reason,
we compare the stellar \mls from a wide range of models here, to assess
the robustness of our conclusions. 

To test the consistency of the different SPS models (and later, 
the effect of different IMFs),
we constructed a sequence of single-metallicity exponential SFH models
with a range of metallicities and exponential $e$-folding timescales.
Then, for each SPS model, we compare the correlation 
between $B - R$ color and stellar \ml in a variety of passbands.

We show the effect of different SPS models in Fig.\,\ref{fig:modelcomp1}
and in Table \ref{tab:mlsps} in Appendix\,\ref{sec:app}.
We adopt a Salpeter IMF, and show the color--\ml relation for 
solar metallicity $\tau$ models in the $B$ band (thin lines) and 
$K$ band (thick lines).  We show four SPS models:
\citet[solid]{bruzual00} models, 
\citet[dotted]{kodama97} models, \citet[dashed]{schulz00} models
and the updated {\sc p\'egase} models
of \citet[long dashed]{fioc00}.  

For all models we find very similar slopes and zero points for the
color--\ml relation (to within 0.1 dex in \ml;
Fig.\,\ref{fig:modelcomp1}).  This also holds true for other passband
combinations and metallicities.  The only exception to this result is
the \citet{schulz00} model, which have an unusually bright asymptotic
giant branch which produces very red optical--near-IR colors for solar
metallicity stellar populations.  The solar metallicity \citet{schulz00}
model gives normal $B$ band stellar \mls but very low $K$ band stellar
\mlsn, compared to the other solar metallicity models.  Essentially, this
means that the \citet{schulz00} solar metallicity model $B - K$ colors
are redder than the other SPS models we compare to (and, indeed, most
of the luminous
spiral galaxies in our observational sample).  This poses a problem,
however, as at a given optical-optical color (e.g.\ $B - R$) the
optical--near-IR colors (e.g.\ $B - K$) of the solar metallicity
\citet{schulz00} models are far too red to explain observed galaxy
colors, whereas the other models do reproduce the observed colors.  In
order to match observed spiral galaxy optical--optical and
optical--near-IR colors simultaneously, 1/3 solar metallicity
\citet{schulz00} models must be adopted.  We plot these models in
Fig.\,\ref{fig:modelcomp1}: these models have stellar \mls much closer
to other models' solar metallicity stellar \mlsn. 
This slight model mismatch is actually quite useful: 
it demonstrates that even with substantial model differences, 
the stellar \ml at a given optical--near-IR color is robust to model 
differences.

\begin{figure}[tb]
\epsfxsize=\linewidth
\epsfbox[20 145 562 668]{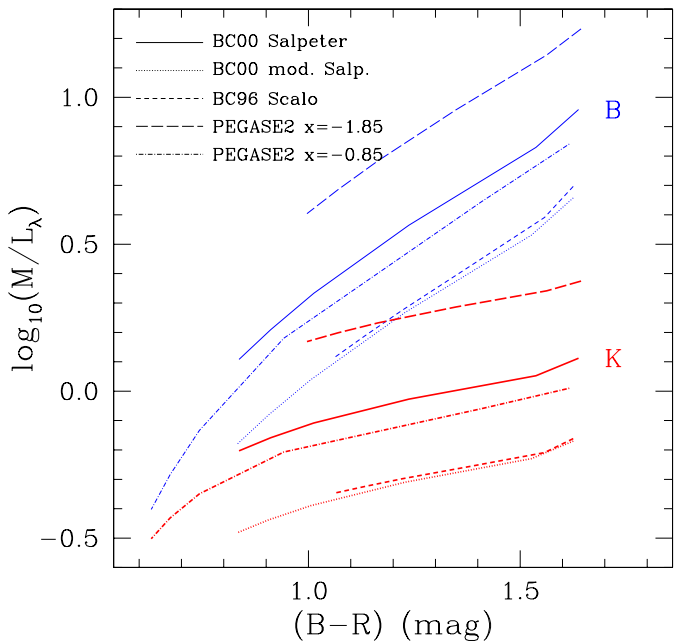}
\caption{\label{fig:modelcomp2}
Comparison of the color--\ml relation for a sequence of exponentially
declining star formation rate models of age 12\,Gyr
using a variety of IMFs.  Again, the thin lines are
for \mln$_B$, the thicker lines are for \mln$_K$. 
The different models and IMFs used are: 
\citet{bruzual00} models with a Salpeter $x=-1.35$ IMF (solid), 
a Salpeter IMF with $x=0$ below 0.6M$_{\sun}$ (dotted), and 
\citet{scalo86} IMF (dashed);
and the updated {\sc p\'egase} models
of \citet{fioc00} with a steeper $x=-1.85$ IMF (long dashed) and a
flatter $x=-0.85$ IMF (dot-dashed).  All models have solar
metallicity.\vspace{-0.2cm} }
\end{figure}

We now test the effect of different IMFs in Fig.\,\ref{fig:modelcomp2}.
We try out a wide range of IMFs for both the \citet{bruzual00} 
and {\sc p\'egase} models:
\citet{bruzual00} models with a Salpeter IMF (with a logarithmic slope 
$x=-1.35$; solid), 
a Salpeter IMF modified to have a flat $x=0$ slope below 
0.6M$_{\sun}$ (dotted), and \citet{scalo86} IMF (dashed);
and the updated {\sc p\'egase} models
of \citet{fioc00} with a steeper $x=-1.85$ IMF (long dashed) and a
flatter $x=-0.85$ IMF (dot-dashed).  All models have solar
metallicity.  The slopes of the color--\ml correlations are 
independent of IMF: only the zero-point is affected by the 
choice of IMF.  The color range is also slightly affected by 
the IMF choice (especially the upper end of the IMF), as the 
range in models is from a single burst at the red end to constant star
formation rate for 12 Gyr at the blue end.  The sensitivity
of the zero point of the color--\ml correlation to the IMF is 
due entirely to differences in the numbers of low mass stars in 
each IMF.  These low mass stars significantly change the total mass of the 
stellar population, but hardly change the overall color 
and luminosity of the system
(which is dominated by the more massive stars).  This justifies 
the scaling of the Salpeter IMF that we have done to 
bring the stellar \mls of the Salpeter IMF into line with
the maximum disk constraints in \S \ref{sec:rotn}: this scaling
has the same effect as a flattening of the low mass end of the IMF.

We therefore conclude that our choice of stellar population synthesis
model does not significantly affect our conclusions: in particular,
the {\it relative} trend in stellar \ml with color is preserved in all
of the models which we examined.  However, the model IMF does make a
significant difference: while the IMF leaves the slope of the
color--\ml correlation and the colors 
relatively unaffected, the IMF strongly affects
the overall normalization of the stellar \mln.

%figure:modelcomp1
\begin{figure}[tb]
\epsfxsize=\linewidth
\epsfbox[20 145 562 668]{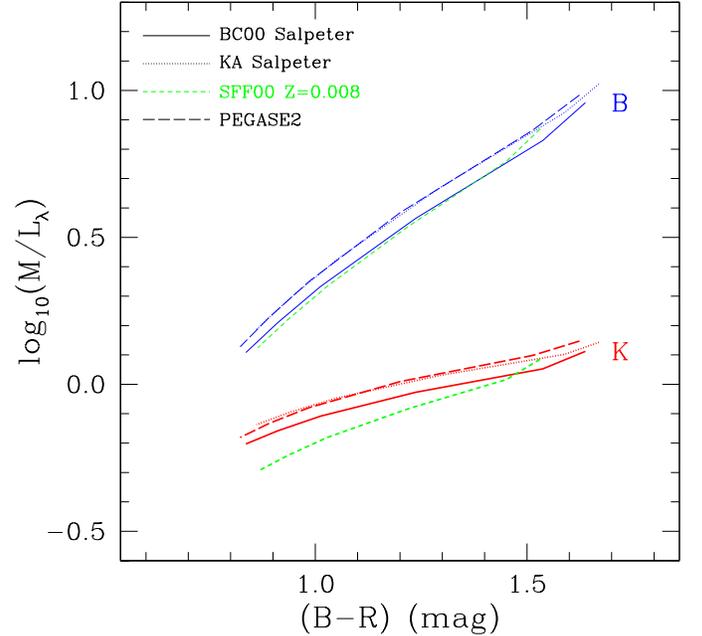}
\caption{\label{fig:modelcomp1}
Comparison of the color--\ml relation for a sequence of exponentially
declining star formation rate models of age 12\,Gyr
using a variety of SPS models. The red end of
the lines represent a short burst of star formation, the blue end
represents a constant star formation rate model. The thin lines are
for \mln$_B$, the thicker lines are for \mln$_K$. 
The different models used
are: \citet[solid]{bruzual00}, 
\citet[dotted]{kodama97}, \citet[dashed]{schulz00} 
and updated {\sc p\'egase} models
of \citet[long dashed]{fioc00} all with a Salpeter IMF.  All models have solar
metallicity except for the \citet{schulz00} models which 
have 1/3 solar metallicity (see text for more details).\vspace{-0.2cm} }
\end{figure}

%figure:modelcomp2

\subsection{Galaxy evolution uncertainties} \label{subsec:galevo}

In this section, we examine the uncertainties stemming from 
differences in galaxy evolution prescriptions.
We have already examined the properties of six different galaxy 
evolution models in \S \ref{sec:ml} and Appendix \ref{sec:app}.  
We found that there was little
difference between the behaviors of the closed box, infall, outflow, 
dynamical time, mass-dependent formation epoch and 
mass-dependent formation epoch with bursts 
models.  In particular, the trends in stellar \ml with color, and their 
zero-points, were remarkably robust to a variety of different effects, 
including low-level bursts in the SFH.  In addition, we have tested
the effects of changing the age of galaxies at
the present day from 12 Gyr: age changes of $\pm 3$ Gyr produce
changes in model stellar \ml at a given color of only $\pm 0.05$ dex.

%figure:starburst
\begin{figure*}[tb]
\hbox{%
\epsfxsize=8.8cm
\epsfbox[20 145 580 668]{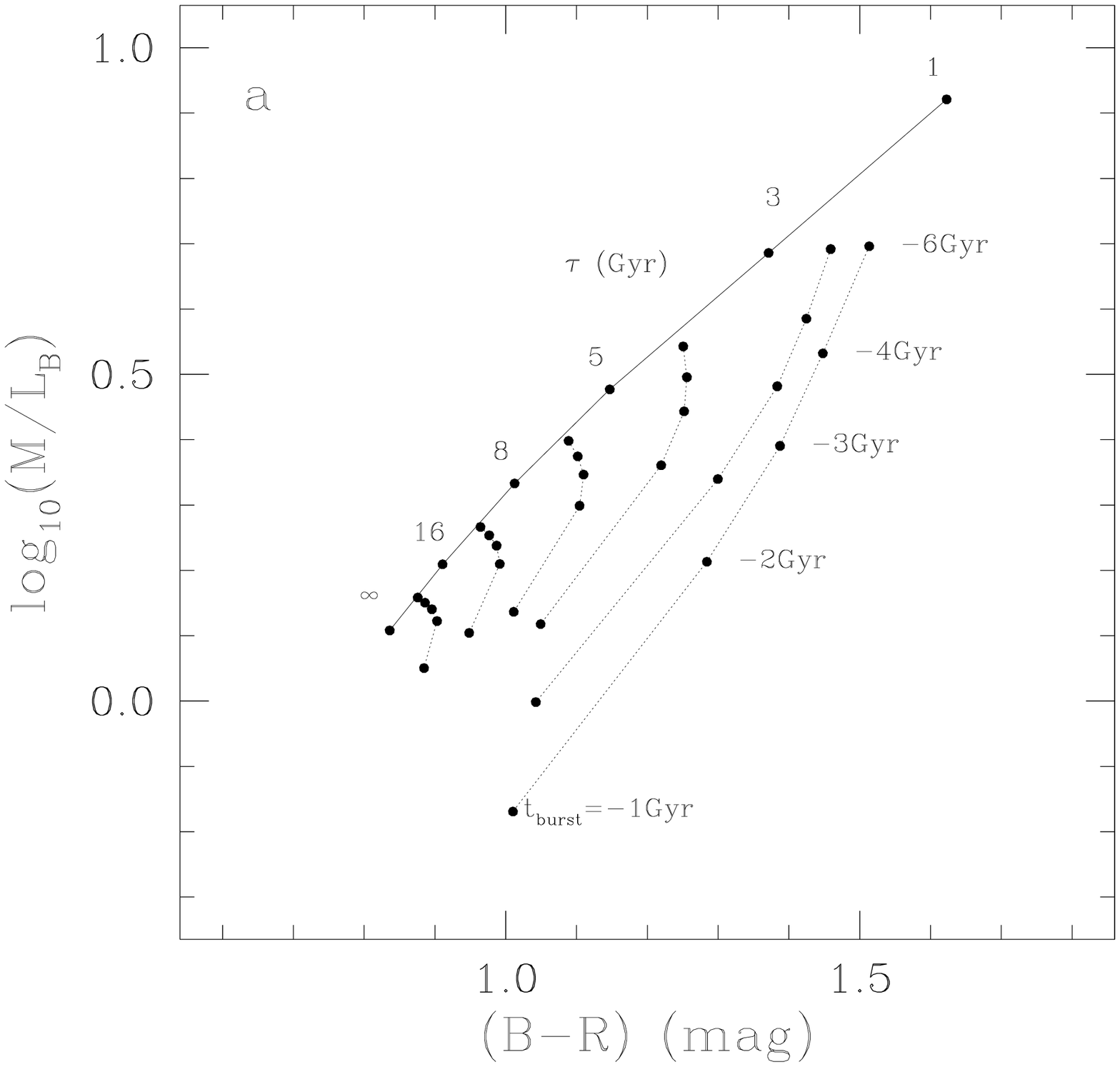}
\hfill
\epsfxsize=8.8cm
\epsfbox[20 145 580 668]{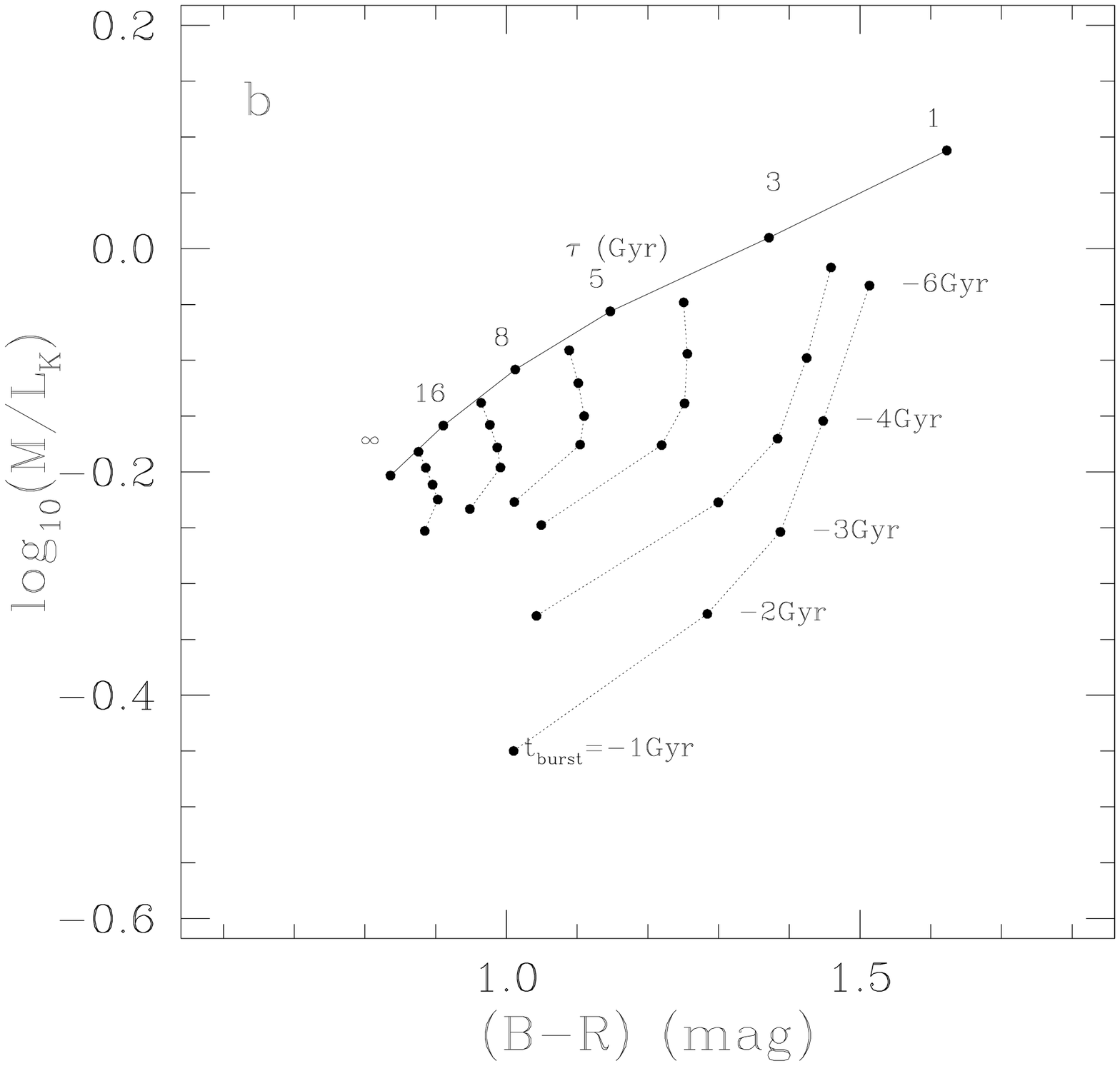}
%\epsfbox[20 145 562 668]{f5b.ps}
}
\caption{\label{fig:starburst}
Color--\ml relations in $B$ (a) and $K$ (b) for a sequence of exponential
declining star formation rate solar metallicity models of age 12\,Gyr
with 10\% mass fraction added in 0.5\,Gyr star bursts. The solid line
connects the exponential SFH models with different $e$-folding times
scales $\tau$. The dotted lines connect models of the same $\tau$
value, but with added star bursts occuring 1, 2, 3, 4 or 6\,Gyr ago.\vspace{-0.2cm}
}
\end{figure*}

One important issue is the effects of larger bursts: do galaxies with
a recent or ongoing burst of star formation have stellar \mls which 
vary considerably from the stellar \mls of galaxies with more 
quiescent star formation but the same colors?  We tested this 
case by adding a star burst with 0.5\, Gyr duration to a range of exponential
SFH models with a mass fraction of 10\%
of the total stellar mass formed over the lifetime of the galaxy.  
We viewed these models at a range of times after the burst,
between 1 and 6 Gyr.  A number of points are apparent from inspection of 
Fig.\ \ref{fig:starburst}.  Firstly, the effects of a 10\%
burst of star formation are much larger for red earlier-type galaxies
than for blue, later-type galaxies.  This stems from the larger fractional
contribution of the young stars to the total {\it luminosity} in redder
galaxies.  Secondly, maximum offsets from the color--stellar \ml correlation
are expected to be $\sim$0.5 dex in $B$, and $\sim$0.3 dex in $K$ band.
Thirdly, bursts of star formation bias the stellar \ml to lower 
values at a given color.  Finally, large effects are only visible
for a period of $\sim$1 Gyr for bluer underlying stellar populations, but are
visible for much longer ($\sim$5 Gyr) for redder underlying stellar 
populations.  

This at first sight seems discouraging: in particular, the sensitivity
of the stellar \ml of redder underlying populations to a burst of 
star formation several Gyr ago implies significant scatter in the 
stellar \mls of redder galaxies.  This is part of our motivation
for choosing a model with bursts of star formation as our default: 
with a model which incorporates bursts of star formation, we 
can account for the lower stellar \mls of redder galaxies with even 
modest amounts of bursty star formation several Gyr ago (compare
panels c and d of 
Fig.\ \ref{fig:amod} in Appendix \ref{sec:app}).  However, we can take some
comfort from the fact that our use of a 10\% burst 
is very conservative: recent bursts of star formation that large are
unlikely, and are likely to be selected against in sample selection 
(by e.g.\ selecting for undisturbed and symmetric galaxies). 
Indeed, even if morphological selection does not filter out these galaxies, 
galaxies with such large bursts are expected to lie 
off of the TF relation (because their luminosities will have been 
considerably boosted by the starburst), and so may be selected against
for this reason. 

As a check, 
we have also examined the trends in stellar \ml with color using
disk-dominated non-satellite galaxies from the heirarchical models of
\citet{cole00}.  These models include the effects of halo formation
and merging, gas cooling, star formation, feedback and dust (but use
the same SPS models as we adopt for this
paper, with a \citet{kennicutt83} IMF and a 38\% brown dwarf fraction), 
and therefore offer a completely independent assessment of the
effects of galaxy evolution prescriptions on the stellar \mls of
galaxies.  The trend in their disk galaxy model 
stellar \mls with color is almost identical to
those of the simpler models (in particular to the 
mass-dependent formation epoch with bursts model),
albeit with more scatter due to the strongly
irregular SFH (panel f of Fig.\ \ref{fig:amod} of Appendix \ref{sec:app}).  
The key to the relatively modest scatter in model stellar \ml
with color in their models can be linked to the morphological
transformations which accompany large mergers. Mergers large enough to 
produce large starbursts with large decreases in stellar \mln, 
are large enough to transform
a disk-dominated into a spheroidal galaxy: these galaxies would not be 
included in any disk-dominated sample of galaxies.

We therefore conclude that choosing a different galaxy
evolution prescription would not significantly affect
the trends in model stellar \ml with color presented in this paper.
Large bursts of recent star formation can lower the 
stellar \ml at a given color by up to a factor of three, however
galaxies with a large amount of recent star formation are unlikely
to feature heavily in a spiral galaxy sample.  The 
lower-level bursts more typical of disk-dominated spiral galaxies add
only modest amounts of scatter to the color--stellar \ml correlation
and are accounted for by our default model.

\subsection{Dust} \label{sec:dust}

Another potential concern is dust: dust simultaneously reddens
and dims a stellar population, changing both axes in the 
correlation between color and stellar \mln.  We address this 
problem in panel d of Fig.\,\ref{fig:ml}, where we show dust extinction
vectors for the dust correction of \citet{tully98}
in $B$ and $K$ band.  
Dust extinction vectors for screen and Triplex models
\citep{ddp} are similar in direction to this vector.
The dust vector shown in Fig.\,\ref{fig:ml} represents a large effect:
it is the correction
to face-on suffered by a Milky Way-type galaxy viewed at an inclination
angle of 80 degrees.  For most galaxies the effects of extinction 
will be much smaller.  It is clear that dust is a 
second order effect for estimating stellar \mls in this 
way.  Dust extinguishes light from the stellar population, making
it dimmer.  However, dust also reddens the stellar population, making 
it appear to have a somewhat larger stellar \mln.  To first order, 
these effects cancel out, leaving a dust-reddened galaxy on the 
same color--stellar \ml correlation.  There is a possibility of
overpredicting (underpredicting) the stellar \ml (thus the stellar mass) if 
not enough (too much) reddening correction is applied, as the reddening
effect is larger than the extinction effect.  However, even for 
the large extinction error illustrated here, the effect is of order 0.1--0.2
dex.  This error is comparable to the errors from uncertainties
in stellar population synthesis modelling and galaxy evolution
prescriptions.  However, this may not apply on a pixel-to-pixel
level:  some small regions of spiral galaxies may be optically thick 
in the optical, which completely obscures the light without producing 
any extra reddening \citep[e.g.][]{witt92}.  
Therefore, smaller scale applications of
this color-based stellar \ml technique must be wary of the effects of 
dust.

\subsection{Summary}

The color--stellar \ml correlation is robust in a relative
sense (both within a passband and between passbands), {\it provided there is 
no systematic change in IMF with galaxy type}.  Model uncertainties, 
galaxy evolution prescription uncertainties, small bursts of star formation 
and dust uncertainties
are all of order 0.1--0.2 dex or 
less.  Large bursts of recent star formation may produce quite
a large effect, depending on when they happen and on the properties of
the underlying older stellar population.  However, large bursts
are unlikely to be common \citep[at least at the present day: at higher 
redshift this need not be the case; e.g.][]{brinchmann00}.
The IMF remains the largest uncertainty:
assuming no trend in IMF with galaxy type, the range of IMFs presented
in the literature causes uncertainty
in the absolute normalization of the stellar \mls of at least a factor
of two.  We address this normalization in the next section.

%----------------------------------------------------------------------
\section{ROTATION CURVES AND THE NORMALIZATION OF THE STELLAR M/L} 
  \label{sec:rotn}

We demonstrated that the model color--stellar \ml
correlation is robust in a relative sense, but has uncertain
overall normalization.  For many applications, this is perfectly
acceptable.  For example, it is quite possible to investigate the 
{\it slope} of the stellar mass TF relation, or estimate the 
trend in stellar \ml as a function of galaxy radius for 
rotation curve fitting, without knowing the absolute normalization of
the overall stellar \mln.  However, for some applications, e.g.\
for understanding the slope of the {\it baryonic} TF relation, or 
in constraining the shape of dark matter haloes, it is important
to understand both the relative trend of stellar \ml with color
and the absolute normalization.  
The previous section showed that the 
question of the absolute normalization of the stellar \ml 
essentially boils down to one issue: the stellar IMF.  
To first order, the amount of stellar light produced by 
observationally plausible IMFs is rather similar; however, the 
slope of the IMF, especially at the low-mass end, changes the 
overall stellar mass considerably.  

We cannot address this problem fully, short of counting all 
of the stars in spiral galaxies directly.  However, we can 
provide some constraints.  The rotation curves of spiral galaxies
have contributions from the stellar mass, gas mass and dark matter.
The relative contributions of each are difficult to estimate 
directly.  However, interesting constraints can be derived by 
assuming that the mass of the stellar disk 
makes the maximum possible contribution
to the rotation velocity: this is 
the maximum disk hypothesis \citep[e.g.][]{vanalbada86}.
Fitting a maximal stellar disk to a rotation curve provides the 
maximum possible stellar \mln, thus providing a firm upper limit 
to the stellar \mls that we have constructed in the model.

%figure:maxdisk
\begin{figure}[tbh]
\epsfxsize=\linewidth
\epsfbox[187 363 407 591]{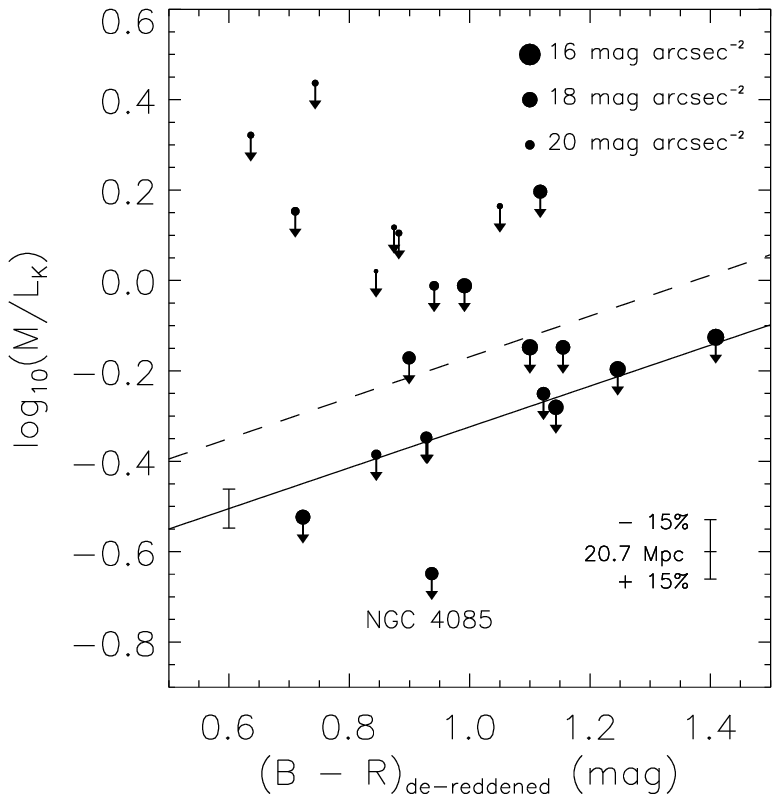}
\caption{\label{fig:maxdisk}  
Observed $K$ band maximum disk stellar \mls against de-reddened $B - R$ color.
The data are from $K$ band imaging and \hi rotation curves
from \protect\citet[Chapter 6]{verheijen97}, rescaled to 
a distance of 20.7 Mpc \protect\citep{sakai00}: the effect on the maximum 
disk \mls of a $\pm 15$\% Ursa Major Cluster distance error is 
also shown.  Overplotted is the least-squares
fit to the correlation between color and stellar \ml for the
formation epoch with bursts model assuming a Salpeter (dashed line)
and a scaled-down Salpeter IMF (solid line).  We also show the 
RMS spread of the formation epoch with bursts model
around the color--\ml relation 
on the solid line as an error bar.  NGC\,4085
is highlighted: it has a poorly resolved rotation curve, which biases
the maximum disk \ml downwards.  Symbol size is coded by inclination-corrected
$K$ band central surface brightness.\vspace{-0.2cm}
}
\end{figure}

We have examined the $K$ band maximum disk stellar \mls of the Ursa
Major Cluster sample
of \citet[Chapter 6]{verheijen97}, rescaled to 
the HST Key Project distance 
of 20.7 Mpc \protect\citep{sakai00} to place constraints
on the normalization of the stellar \mlsn.  This value is 
consistent (bearing in mind $\ga$10\% systematic uncertainties)
with the distance derived from a different analysis of 
the Cepheid-calibrated TF relation \citep[18.6 Mpc;][]{tully00}
and the brightness of a Type 1a supernova in NGC 3992 
which was consistent with a distance of 24$\pm$5 Mpc \citep{parodi00}.
$K$ band was 
adopted as we have shown above that using the $K$ band results in the most
robust stellar \ml estimation.  We consider
the maximum stellar \ml given by either the pseudo-isothermal 
or Hernquist halo fit.  In Fig.\,\ref{fig:maxdisk} 
we plot this $K$ band maximum
disk stellar \ml against the $B - R$ color
of the galaxy, de-reddened assuming dust extinction following
\citet[see also \S \ref{sec:tf}]{tully98}. These are the
dynamical upper limits for the stellar \mls of 
these galaxies, hence the upper limit signs.

NGC\,4085 is highlighted: this nearly edge-on galaxy was observed with a
beam the size of its minor axis diameter, resulting in the worst
case scenario for beam smearing \citep[e.g.][]{vdb00}. Consequently, 
it has a poorly resolved rotation curve, 
which biases the maximum disk \ml downwards.  We ignore the stellar
\ml estimate for NGC\,4085 
in the following discussion, although clearly a better resolved 
rotation curve would be useful.

The main point of this plot is that our SPS-based model stellar \mls should 
be the same as or lower than all of the observed maximum disk stellar
\mlsn.  We make the explicit assumption here that the lower envelope
of the observed maximum disk stellar \mls is the meaningful constraint
(again, we neglect NGC\,4085 due to beam smearing).
Galaxies with maximum disk \mls significantly above this envelope
are interpreted as galaxies with significant dark matter within the 
optical radius of the galaxy: these galaxies are sub-maximal.
This interpretation is supported by the surface brightnesses of the 
sub-maximal disks: they are all fairly low surface brightness.  
Low surface brightness galaxies are thought to
have high maximum disk stellar \mls because they are 
dark matter dominated even in their inner regions 
\citep[e.g.][]{verheijen97,deblok98}.

From Fig.\,\ref{fig:maxdisk}, it is clear that applying our standard
color--stellar \ml relation assuming a Salpeter $x=1.35$ IMF 
normalization over-predicts
the stellar \ml of many of the galaxies (dashed line).  
Motivated by recent IMF determinations which suggest a turn-over
in the IMF at low stellar masses \citep[e.g.][]{kroupa93,larson99,kroupa00}
we scale down the Salpeter IMF masses by a factor of 0.7.  This is equivalent
to a Salpeter IMF $x=1.35$ with a flat $x=0$ slope below 0.35M$_{\sun}$, or 
a \citet{kennicutt83} IMF with a brown dwarf fraction of $\sim$ 40\%.
This scaled IMF results in the solid line
in Fig.\,\ref{fig:maxdisk}.
This IMF is maximal: the stellar \mls can be no larger
than those predicted by a model adopting this IMF, modulo
distance uncertainties.  The maximum disk \mls scale inversely 
with distance: a 15\% error bar for the data points 
is shown, corresponding to a 10\% random and 10\% 
systematic error added in quadrature \citep{sakai00}.
Indeed, the stellar \mls might have to be even somewhat lower: 
all disks may be sub-maximal \citep[e.g.][]{bottema97,courteau99}, the 
$K$ band maximum disk stellar \ml has not been corrected for 
the effects of dust extinction, and the mass locked up in 
molecular hydrogen has not been accounted for in 
these rotation curve decompositions.  
On the other hand, the \hi rotation curves are all to some extent affected
by at least small amounts of beam smearing (which would work to 
lower the maximum disk stellar \ml estimate): the upshot is that there
is some scope for moving the stellar \mls only slightly upwards, and
there is much scope for moving the stellar \mls substantially downwards, 
lending credibility to the idea that our scaled Salpeter IMF is maximal.

One remarkable point is that, modulo the modest sample size, 
the slope of the lower envelope of the observational maximum
disk stellar \mls is accurately described by the predicted
trend in $K$ band stellar \ml with $B - R$ color.  The zero-point
of the model has been constrained to match the data; however, 
there was no {\it a priori} reason that the slope of the observational
color--stellar \ml relation needed to match the predictions of the model.
This is remarkable for a few reasons.  Firstly, it puts our proposition
that the stellar \ml is primarily a function of color, varying a
factor of two in the $K$ band between the reddest and bluest galaxies,
on a more empirical footing.  Secondly, it suggests that galaxies close to
maximum disk have very similar IMFs, as
strong IMF variations with galaxy color should be easily visible
in this plot.  In fact, the scatter of the observational lower envelope 
around the predicted line is consistent with the predicted model scatter due
to differences in SFH at a given color, leaving {\it no} freedom for
random galaxy-to-galaxy IMF variations.  Finally, it implies that the
galaxies closest to the observed limit (high surface brightness galaxies in
general), are probably close to maximum disk, because the adopted IMF
already gives a reasonably low \ml zero-point, compared to other IMFs.
At least the \mls must be scaled to a relatively well-defined
maximum disk fraction (to better than $\la$ 0.1 dex, or 25\%), 
which carries with it strong implications for scenarios of 
galaxy formation and evolution. 

The above considerations have led to our preferred stellar \ml model:
we require that the model reproduces trends in color-based stellar
ages and metallicities \citep[ {and} \S 2]{papiii}, properly accounts for the 
decrease in the color--stellar \ml slope caused by modest bursts
of star formation (\S \ref{sec:unc}), 
and has an IMF consistent with maximum disk constraints (this section).
These requirements are met by the mass-dependent formation epoch with bursts
model, adopting a scaled Salpeter IMF (Fig.\ \ref{fig:ml}).
We present least-squares fits to the color--stellar \ml trend
in Table \ref{tab:mladopt}.  These fits can be used to estimate
a stellar \ml for a spiral galaxy stellar 
population of a given color, calibrated to maximum disk. 
If {\it all} (even very high surface brightness)
galaxy disks are sub-maximal, the model fits should be scaled
down by an appropriate, constant factor.
The fits to our preferred model 
reproduce the color--\ml trends
of the other models with this IMF to better than 0.1 dex (Fig.\,\ref{fig:amod}
and Table \ref{tab:mlcol} in Appendix \ref{sec:app}).
These fits are illustrated in panel d of Fig.\,\ref{fig:ml}, 
and in Figs.\,\ref{fig:acol} and \ref{fig:amod} by the
straight lines.  The full models, and fits of stellar \ml against
colors not considered in this paper are available from 
the authors.  In particular, fits of the stellar \ml with 
colors in the Sloan system will become available when the final 
bandpasses are defined.

\begin{table*}
\begin{footnotesize}
\begin{center}
\caption{Stellar M/L as a function of color for the formation 
	epoch model with bursts, adopting a scaled Salpeter IMF
	{\label{tab:mladopt}}}
\begin{tabular}{lcccccccccccccc}
\tableline
\tableline
{Color} & {$a_B$} & {$b_B$} 
& {$a_V$} & {$b_V$} 
& {$a_R$} & {$b_R$} 
& {$a_I$} & {$b_I$}  
& {$a_J$} & {$b_J$}  
& {$a_H$} & {$b_H$}  
& {$a_K$} & {$b_K$} \\
\tableline
$B - V$ & $-$0.994 & 1.804 & $-$0.734 & 1.404 & $-$0.660 & 1.222 & $-$0.627 & 1.075 & $-$0.621 & 0.794 & $-$0.663 & 0.704 & $-$0.692 & 0.652 \\ 
$B - R$ & $-$1.224 & 1.251 & $-$0.916 & 0.976 & $-$0.820 & 0.851 & $-$0.768 & 0.748 & $-$0.724 & 0.552 & $-$0.754 & 0.489 & $-$0.776 & 0.452 \\ 
$V - I$ & $-$1.919 & 2.214 & $-$1.476 & 1.747 & $-$1.314 & 1.528 & $-$1.204 & 1.347 & $-$1.040 & 0.987 & $-$1.030 & 0.870 & $-$1.027 & 0.800 \\ 
$V - J$ & $-$1.903 & 1.138 & $-$1.477 & 0.905 & $-$1.319 & 0.794 & $-$1.209 & 0.700 & $-$1.029 & 0.505 & $-$1.014 & 0.442 & $-$1.005 & 0.402 \\ 
$V - H$ & $-$2.181 & 0.978 & $-$1.700 & 0.779 & $-$1.515 & 0.684 & $-$1.383 & 0.603 & $-$1.151 & 0.434 & $-$1.120 & 0.379 & $-$1.100 & 0.345 \\ 
$V - K$ & $-$2.156 & 0.895 & $-$1.683 & 0.714 & $-$1.501 & 0.627 & $-$1.370 & 0.553 & $-$1.139 & 0.396 & $-$1.108 & 0.346 & $-$1.087 & 0.314 \\ 
\tableline \\
\end{tabular} 
\end{center}
\vspace{-0.2cm} Note. --- $\log_{10}({\rm M/L}) = a_{\lambda} + b_{\lambda} {\rm Color} $ \\ 
Note that the stellar \ml values can 
be estimated for any combination of the above colors by a simple linear
combination of the above fits.  Note also that if {\it all} (even very high
surface brightness) disks are
sub-maximal the above zero points should be modified by subtracting 
a constant from the above relations. \vspace{-0.2cm}
\end{footnotesize}
\end{table*}

%----------------------------------------------------------------------
\section{THE TULLY-FISHER RELATION} \label{sec:tf}

Having established that galaxy evolution models make robust
predictions of a correlation between optical colors and stellar \mlsn,
we will now investigate the implications for the TF relation.
The TF relation relates
the dynamical mass of a galaxy to its luminosity, thus providing
a stringent test of theories of galaxy formation 
and evolution \citep[e.g.\ ][]{cole00,navarro00b,vdbosch00}.
However, its power as a test of theories is limited by its passband
dependent slope \citep[this assumes linearity of the TF relation, 
which seems a reasonable assumption over much of the 
TF relation, although the TF relation may be nonlinear at
low galaxy masses: e.g.][]{matthews98,mcgaugh00}.
The slope of the TF relation varies from around $L \propto V^3$ in 
the blue to $L \propto V^4$ in the near-IR.  Depending 
on which passband a theory compares its TF relation to,
it is possible to have a favorable
comparison with one particular TF relation but provide a poor match
to a TF relation at a different wavelength.  There are, of course, 
more complex models which include realistic stellar population 
prescriptions and may be able to reproduce the TF relations at
many wavelengths \citep[e.g.\ ][]{heavens99,cole00}; however, it would clearly
be useful to be able to compare the models with one, unique, 
passband-independent TF relation.

In this section, we apply the trends in stellar \ml with spiral
galaxy color described in Table \ref{tab:mladopt} 
to the TF relation data of \citet{verheijen97}
with a dual aim.  Firstly, we wish to test 
the stellar \mls derived
in \S\,\ref{sec:ml} to check if the stellar masses derived
from different passbands give consistent results.  Secondly, we wish
to find out if there is a single, passband-independent TF relation, and
if so, what is its slope (assuming a linear TF relation)?  
The identification of a single, 
passband-independent TF relation will allow even simplistic models
to compare meaningfully with observations without having to construct
a complex and realistic SFH model.

\subsection{The data}

%figure:tf1
\begin{figure}[tbh]
\epsfxsize=\linewidth
\epsfbox{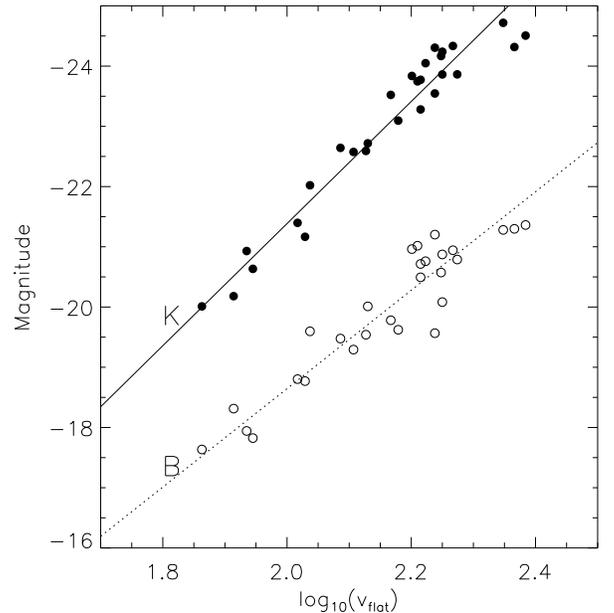}
\caption{\label{fig:tf1} The Tully-Fisher relation in
$B$ and $K$ passbands.  
Solid and open circles denote the data of \protect\citet{verheijen97}
in $K$ and $B$ bands respectively, 
corrected using Tully et al.'s \protect\citeyearpar{tully98}
mass-dependent dust corrections.  The lines
denote the least squares bisector fits
\protect\citep{isobe90} to the mass-dependent
dust corrected TF relations. \vspace{-0.2cm}}
\end{figure}

Here, we use the TF data obtained by \citet{verheijen97} of the  Ursa
Major Cluster.  The Ursa Major Cluster is a nearby
\citep[HST Key Project distance $D = 20.7$ Mpc;][]{sakai00}, 
poor cluster rich in spiral galaxies.  The Verheijen data set
is particularly suitable for our purposes because it
provides accurate magnitudes in $B$, $R$, $I$ and $K'$ 
and has accurate rotation velocities from well-resolved \hi aperture synthesis
rotation curves.  We here consider only the rotation velocity at the flat
part of the rotation curve ($v_{\rm flat}$): \citet{verheijen97} concludes
that use of this rotation velocity minimizes the scatter of
the TF relation.  Furthermore, the rotation velocity at the flat part of 
the rotation curve is a `clean' observational quantity at a reasonably 
well-defined radial range.  The \hi linewidth is a much more ill-defined
quantity, resulting from the interplay of the rotation curve and global 
\hi distribution (even neglecting the influence of warps, asymmetries, 
kinematic irregularities, and gaseous velocity dispersion).  
Thus, while the use of linewidth-based TF relations
for distance estimation purposes is perfectly valid, the use of linewidths
for constructing the {\it intrinsic} TF relation as a test of galaxy evolution
models is far from ideal. Using $v_{\rm flat}$ is much fairer, 
and better reflects the true relationship between the rotation 
velocity of a galaxy and the stellar populations in that galaxy.

We correct for foreground galactic extinction 
assuming a $B$ band extinction of 0.08 mag \citep{schlegel98}.
We further correct for extinction internal to the galaxy following
\citet{tully98}, who determined a galaxy linewidth
dependent extinction correction by minimizing scatter around the 
color--magnitude and TF relation for a sample of 87 galaxies (although
the Ursa Major Cluster galaxies form part of the dataset defining
the dust correction, meaning that the dust correction we use
was partially derived from the TF relation data we analyze here).
According to this recipe, high mass galaxies have a significant 
extinction correction, whereas low mass galaxies have a negligible 
extinction correction.  We adopt the linewidth-dependent 
version of this correction.
Independent support for a mass-dependent
extinction correction comes from \citet{dejong00}, who use
a sample of nearly 1000 spiral galaxies to find
that high surface brightness (usually luminous) galaxies have optical depths of
the order of 1 in their center, but that low surface brightness 
(usually less luminous) galaxies behave
in a nearly transparent manner.  
For reference, we also tried the 
mass-independent extinction correction applied by 
\citet{verheijen97} based on the method of \citet{tully85}.

The TF relations in $B$ and $K$ bands are shown in
Fig.\,\ref{fig:tf1}, as are the best fit least-squares bisectors
(see also Table \ref{tab:slope}).  Least squares
bisectors \citep{isobe90} 
are the average of the `forwards' and `backwards' fits 
to the TF relation (which have shallower and steeper slopes 
than this fit, respectively), and are particularly suitable for probing the 
intrinsic correlation between two variables. From Table \ref{tab:slope} and 
Fig.\,\ref{fig:tf1}, it is immediately apparent that the 
TF relation is shallower in the bluer passbands than in 
the near-IR, even accounting for magnitude-dependent
dust corrections \citep[e.g.][]{verheijen97,tully98}.  Furthermore, 
the TF relation constructed using mass-independent
dust corrections is shallower than the TF relation 
constructed using Tully et al.'s \citeyearpar{tully98}
mass-dependent dust corrections: the discrepancy worsens
as the passband becomes bluer.  The fact that the TF relation 
steepens at longer wavelengths, even when accounting for mass-dependent
dust corrections, is a clear indication that the stellar \ml
varies with mass in just the way implied by Fig.\,\ref{fig:ml}.

\begin{table*}
\begin{footnotesize}
\begin{center}
\caption{Intercepts and slopes of the TF relations: 
	$L = L_{100} V^{\alpha}$ and $M = M_{100} V^{\alpha}$
	{\label{tab:slope}}} 
\begin{tabular}{lcccccccc}
\tableline
\tableline
{\bf Luminosities} & \multicolumn{2}{c}{$B$} & \multicolumn{2}{c}{$R$} &
\multicolumn{2}{c}{$I$} & \multicolumn{2}{c}{$K$} \\
{Case}     & $\log_{10} L_{100}/L_{\sun}$  & $\alpha$ &
$\log_{10} L_{100}/L_{\sun}$  & $\alpha$ &
$\log_{10} L_{100}/L_{\sun}$  & $\alpha$ &
$\log_{10} L_{100}/L_{\sun}$  & $\alpha$ \\
\tableline
Mass-dep dust  & 9.65 $\pm$ 0.03 & 3.27 $\pm$ 0.17 & 
 9.60 $\pm$ 0.03 & 3.54 $\pm$ 0.16 & 
 9.62 $\pm$ 0.03 & 3.77 $\pm$ 0.17 &
 9.89 $\pm$ 0.03 & 4.06 $\pm$ 0.20 \\
Mass-indep dust & 9.84 $\pm$ 0.03 & 2.76 $\pm$ 0.15 & 
 9.69 $\pm$ 0.03 & 3.18 $\pm$ 0.15 &
 9.68 $\pm$ 0.03 & 3.46 $\pm$ 0.17 &
 9.88 $\pm$ 0.03 & 3.98 $\pm$ 0.20 \\
\tableline
{\bf Masses} & 
\multicolumn{2}{c}{$B$} & \multicolumn{2}{c}{$R$} &
\multicolumn{2}{c}{$I$} & \multicolumn{2}{c}{$K$} \\
{Case}     & $\log_{10} M_{100}/M_{\sun}$  & $\alpha$ &
$\log_{10} M_{100}/M_{\sun}$  & $\alpha$ &
$\log_{10} M_{100}/M_{\sun}$  & $\alpha$ &
$\log_{10} M_{100}/M_{\sun}$  & $\alpha$ \\
\tableline
Stellar mass (MD)& 9.51 $\pm$ 0.04 & 4.34 $\pm$ 0.22 
 & 9.51 $\pm$ 0.04 & 4.34 $\pm$ 0.22 & 
 9.49 $\pm$ 0.04 & 4.49 $\pm$ 0.23 & 
 9.49 $\pm$ 0.04 & 4.51 $\pm$ 0.26 \\
% Fit to everything, 9.50 $\pm$ 0.04, 4.44 $\pm$ 0.23
Stellar mass (MI)& 9.38 $\pm$ 0.04 & 4.33 $\pm$ 0.23 
 & 9.38 $\pm$ 0.04 & 4.33 $\pm$ 0.23 & 
 9.35 $\pm$ 0.04 & 4.49 $\pm$ 0.24 &
 9.37 $\pm$ 0.04 & 4.62 $\pm$ 0.25 \\
% Fit to everything, 9.37 $\pm$ 0.04, 4.47 $\pm$ 0.24
Baryonic mass (MD) & 9.79 $\pm$ 0.04 & 3.45 $\pm$ 0.18 
 & 9.79 $\pm$ 0.04 & 3.45 $\pm$ 0.18 
 & 9.78 $\pm$ 0.04 & 3.55 $\pm$ 0.19 
 & 9.79 $\pm$ 0.04 & 3.51 $\pm$ 0.19 \\
% Fit to everything, 9.79 $\pm$ 0.04, 3.50 $\pm$ 0.19
\tableline \\
\end{tabular}
\end{center}
\vspace{-0.2cm} Note. --- $L_{100}/L_{\sun}$ and $M_{100}/M_{\sun}$ are luminosities 
	and masses in solar units for a galaxy on the TF relation with 
	a $v_{\rm flat}$ of 100\,km\,s$^{-1}$.  Case (MI) uses
	\protect\citet{tully85} mass-independent dust corrections, 
	and  Case (MD) uses
	\protect\citet{tully98} mass-dependent dust corrections.
	Errors denote the uncertainty in the formal fit to the TF relations. \vspace{-0.2cm}
\end{footnotesize}
\end{table*}

%figure:tf2
\begin{figure*}[tbh]
\epsfxsize=\linewidth
\epsfbox{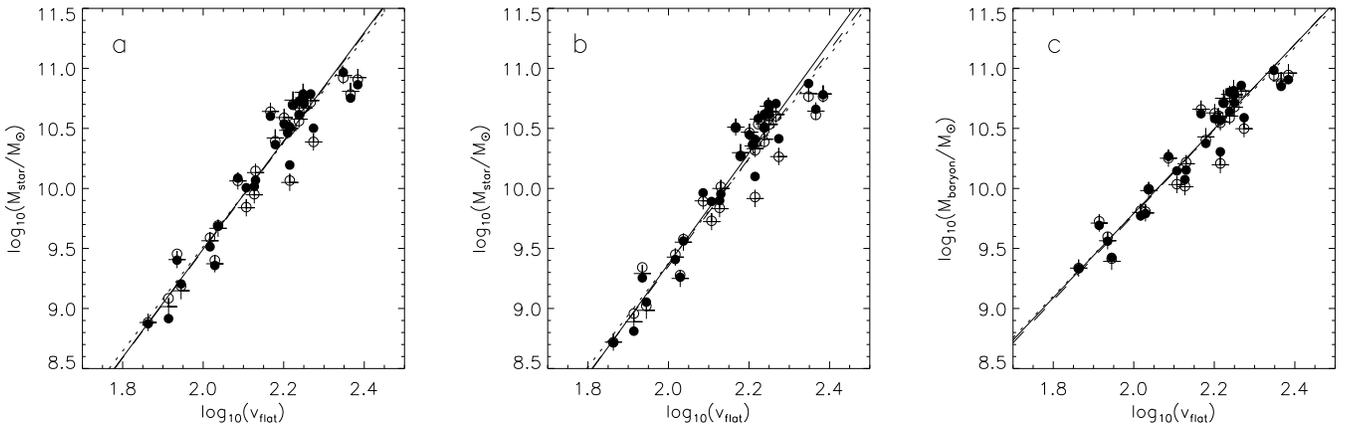}
\caption{\label{fig:tf2} TF relations: stellar mass with mass-dependent
extinction correction (a), stellar mass with 
mass-independent extinction
correction (b) and baryonic TF relation with 
mass-dependent extinction correction (c).  Masses derived from 
$B$ and $R$ data are shown as open circles
(the masses are identical as $B$--$R$ colors
are used to construct the stellar \mls), $I$ band data by crosses, 
and $K$ band data by solid circles.  
Least squares bisector fits to each passband's
TF relations are presented for the $B$ and $R$ data (dotted lines), 
$I$ data (dashed lines) and $K$ band (solid lines). \vspace{-0.2cm} }
\end{figure*}

\subsection{The stellar mass TF relation}

To test this possibility in more detail, we adopt the least-squares
fit to the variation of stellar \ml with $B - R$ color in $B$, $R$, $I$, 
and $K$ passbands for the preferred model (formation epoch model with bursts,
with a scaled Salpeter IMF).   These model relations are used to convert the 
magnitude-dependent dust-corrected
magnitudes into stellar masses, using the dust-corrected $B - R$ 
color as input.  The use of these model relations is suitable:
the TF relation from \citet{verheijen97} is among the tighest in 
the literature, implying a minimal contribution from large starbursts.
The results are shown in Fig.\,\ref{fig:tf2}, panel a.  Stellar masses 
derived from $B$ and $R$ passbands are shown as open circles (the masses
are identical as the $B - R$ color was used to construct the 
\mlsn), the $I$ band by crosses, and the $K$ band by filled circles.  
Least-squares bisector fits of the TF relations are also shown, 
and given in Table \ref{tab:slope}.

From panel a of Fig.\,\ref{fig:tf2} and Table \ref{tab:slope}, it is clear
that by accounting for the variation in stellar \ml with galaxy color we 
have demonstrated that there is one passband-independent stellar
mass TF relation.  The stellar masses determined from $B + R$, $I$ and
$K$ band data for the individual galaxies 
are consistent to within $\sim 10$\%  RMS, powerfully
demonstrating the utility of this technique and confirming 
that the trends suggested by our models are indeed 
consistent with observations.
Furthermore, the stellar mass TF relation ($L \propto V^{4.4 \pm 0.2}$)
is steeper than even the 
$K$ band TF relation ($L \propto V^{4 \pm 0.2}$).
These errors represent only fitting error: errors in IMF and distance
scale do not affect the slope of the stellar mass TF relation, and 
slope errors from adopting fits for different SFH models are $\sim$ 0.2.
In addition, the zero point of the stellar mass TF relation is
proportional to the distance, as we scale to maximum disk (a 
$\pm$15\% distance uncertainty translates into a $\pm$0.06 dex 
zero point shift in the stellar mass TF relation).

The scatter in the stellar mass TF relation is somewhat less than 0.5 mag, 
which is slightly larger than the scatter in the {\it raw} optical 
and near-IR TF relations.  This is an unavoidable disadvantage of this
technique: not only are we making the TF relation steeper (which increases
the magnitude scatter if some of the scatter is caused by 
velocity or distance errors), 
but we are folding in uncertainties from at least two 
different passbands' data into the stellar TF relation.  However, a 
slightly enhanced scatter is a relatively modest price to pay: the true
strength of this type of analysis is in the recovery of a stellar 
and/or baryonic mass TF relation which is passband independent.

An interesting test is to consider the effects of the dust
correction on the recovered stellar mass 
TF relation.  For example, even assuming
that Tully et al.'s \citeyearpar{tully98} dust correction 
is appropriate statistically, the dust correction is unlikely to 
be accurate on a case-by-case basis.  Thus it is important
to test the effects of choosing a different attenuation for the 
galaxy.  We do this by repeating the above analysis using the 
mass-independent dust-corrected TF relation \citep{verheijen97}, the
results of which are shown in panel b of Fig.\,\ref{fig:tf2} 
and Table \ref{tab:slope}.
Comparing the results in Table \ref{tab:slope}, we confirm the conclusion
drawn about reddening in \S \ref{sec:dust}: the 
stellar mass TF estimated using a mass-independent
dust prescription is almost exactly the same
as the mass-dependent dust case.  
A modest offset of $-$0.13 dex is found, which stems from 
a larger blue optical depth in \citet{tully85} 
compared to \citet{tully98}: this produces bluer
de-reddened $B - R$ colors which lead to an overall offset
in stellar mass TF relation without a change in slope.
One interesting implication of this finding is that we cannot
say how much of the decreasing slope of the TF relation with 
decreasing wavelength is due to dust and how much is due to 
stellar \ml differences. We expect the effects to be roughly
comparable, as Tully et al.'s \citeyearpar{tully98} corrections
seem, at least in a statistical sense, quite appropriate.

\subsection{The baryonic mass TF relation}

When we account for the \hi gas fraction to calculate the total known
baryon mass (panel c of Fig.\,\ref{fig:tf2}), we find 
$m_{\rm baryon} \propto V^{3.5 \pm 0.2}$ (using an unweighted least-squares 
bisector fit). 
Since the baryonic TF relation is of significant astrophysical
importance, it is worth discussing the uncertainties in 
the slope we determine above.
We have used an unweighted least-squares bisector:
the slope of forwards and backwards fits are 
$\sim$ 0.15 shallower and steeper repectively.  
There is an uncertainty of $\pm$0.2 or so depending on which model 
is used as the preferred model.  Furthermore, we have not
accounted for the (fairly unconstrained) molecular hydrogen mass fraction:
if molecular hydrogen were included it would probably steepen
the baryonic TF relation slightly \citep{young89}.  On the other
hand, the absolute normalization of the stellar \ml is maximal, 
which implies that the slope stated above is 
as steep as is allowed by maximum disk: for reference, adopting 
a 63\% velocity (40\% mass)
maximal disk following \citet{bottema97} or \citet{courteau99} would make the 
baryonic TF relation slope shallower by 0.5.
Also, we have assumed the HST Key Project 
distance to the Ursa Major 
Cluster \citep{sakai00}.  The stellar masses are proportional
to distance because we scale to maximum disk;
however, the \hi masses are affected by the distance $D^2$.  
\citet{sakai00} estimate around 10\% random 
and 10\% systematic distance uncertainties: the corresponding $\pm$15\% 
total distance error bars lead to slope changes of somewhat less 
than $\mp$0.1.  This suggests that the random and systematic errors
for the baryonic TF relation slope should be $\sim$ 0.2 each.

It should be noted that the scatter in the baryonic and stellar mass
TF relations can place tight constraints on the allowed variations
in IMF at a given rotation velocity.  The scatter in the baryonic TF relation 
is a modest 0.1 dex, and in the stellar mass TF relation, a 
slightly larger 0.13 dex.  Assuming that {\it all} of the error
is due to IMF variations, a FWHM spread of stellar \mls of somewhat
less than a factor of two is allowed at a given rotation velocity.
This is a firm upper limit as 
we do not account for measurement errors in the luminosity, rotation 
velocity, the intrinsic depth of the cluster, non-circular potentials
\citep{franx92}, or the intrinsic spread
in stellar \mls from SFH variations.  Taken together with the 
suggestive tightness of the lower envelope of observational 
maximum disk stellar \mls in Fig.\,\ref{fig:maxdisk} which argues
against large IMF variations at a given color, there is little 
evidence against a universal spiral galaxy IMF.

One interesting comparison that we can perform is with 
the baryonic TF relation of \citet{mcgaugh00}.
They use a constant stellar \ml in each passband to construct
a baryonic TF relation with a slope which is indistinguishable from
4.  They claim that this strongly rules out CDM-like models.
We disagree with their result for the slope by around 2$\sigma$ (even 
including systematic error):  adopting our slope of 3.5 $\pm$ 0.2 
(random) $\pm$ 0.2 (systematic), the case against
the basic relationship $m \propto V_{\rm halo}^3$ 
predicted by simple CDM models is much weaker
\citep[e.g.][]{vdbosch00,navarro00b}.  

This disagreement is at first sight somewhat 
surprising, as accounting for the larger stellar \mls and 
dust extinctions of redder
galaxies would steepen the baryonic TF relation, relative to one constructed
assuming color-independent stellar \mls and dust correction.
However, the difference can be traced to a combination of
three effects.  
Firstly, and most importantly, \citet{mcgaugh00} use values of stellar \ml 
which are around 30--40\% larger 
than ours (at a typical color for a luminous spiral galaxy), 
and assume a distance 25\% shorter 
than the one we adopt.  This accounts for most of the difference in 
baryonic TF relation slope.
Secondly, \citet{mcgaugh00} use linewidths to 
construct their baryonic TF relation.  For a variety of reasons
outlined earlier, we chose to use the more physically-motivated
rotation velocities at the flat part 
of the rotation curve: this leads to a shallower TF relation by perhaps
as much as 0.2 in terms of the slope
\citep[Chapter 5, his Table 7]{verheijen97}.  Finally, we lack galaxies 
with rotation velocities much lower than 80 km\,s$^{-1}$: 
at present, there is no sample of low mass galaxies 
with sufficiently accurate rotation velocities and photometry
to construct accurate $v_{\rm flat}$ and stellar mass estimates.
The inclusion of low mass galaxies may steepen the TF relation, 
or indicate that at low masses the TF relation is non-linear
\cite[e.g.][]{matthews98}.

%----------------------------------------------------------------------
\section{CONCLUSIONS} \label{sec:conc}

Under the assumption of a universal spiral galaxy IMF, 
we have used stellar population synthesis models in conjunction with 
simplified spiral galaxy evolution models to argue that there
are substantial variations in stellar \ml in optical and 
near-IR passbands, and that these \ml variations are strongly
correlated with stellar population
colors.  The variations in stellar \ml also correlate with other
galaxy properties (albeit with more scatter) such that, on average,
low surface brightness,
high gas fraction and low luminosity galaxies have lower
stellar \mls than high surface brightness, low gas fraction, bright
galaxies.  The changes in stellar \mls over a plausible range of 
galaxy parameters amount to a factor of about 7 in $B$, 3 in 
$I$, and 2 in $K$ band.  In addition, because the 
central regions of galaxies are often redder than their outer
regions, the inner regions of galaxies are likely to have larger 
stellar \mls than the outer regions of galaxies.

This strong correlation between color and stellar \ml is robust
to uncertainties in stellar population and galaxy evolution modeling, 
including the effects of modest bursts of recent star formation.
Larger bursts, which are correspondingly more rare and are typically
selected against in spiral galaxy studies (as evidenced by the modest
scatter in our TF relation), may depress the stellar 
\ml from our expectations by up to 0.5 dex, at most.
In addition, because dust both dims and reddens the light from 
galaxies, uncertainties in the exact amount of dust 
do not significantly affect the stellar mass
estimate for a given galaxy.  The stellar IMF remains the primary uncertainty, 
implying that these trends are relative in a robust sense, but 
the absolute normalization is somewhat uncertain.

We analyzed observed $K$ band maximum disk stellar \mls from 
\citet{verheijen97} to place the relative stellar \mls which we estimate
in this paper on the maximum allowed scale.  We find that a Salpeter IMF
is ruled out by this analysis if the Ursa Major Cluster is placed
at the HST Key Project distance of 20.7 Mpc, and that modification of
the low-mass end of the IMF is required.  
We find that the observed maximum disk 
stellar \mls follow the trend suggested by the models, which
lends independent support for our models, implies that a fraction 
of high surface brightness galaxies are reasonably close to maximum disk, and
suggests a universal spiral galaxy stellar IMF.

We apply these maximum disk-scaled trends in stellar \ml with 
galaxy color to investigate the underlying nature of the 
TF relation.  We find that, using mass-dependent dust 
extinction corrections and the color-dependent stellar \mls
it is possible to estimate stellar masses from different passbands
which are consistent at better than the 10\% level.  
The slope of the stellar mass TF relation of the Ursa
Major Cluster sample is 4.4 $\pm$ 0.2, 
using an least-squares bisector fit.
Including the contribution from the \hi mass, 
we find that the slope of the baryonic 
TF relation of the Ursa Major Cluster 
is 3.5 $\pm$ 0.2 (random) $\pm$ 0.2 (systematic), 
adopting an unweighted bisector fit.  The slope 
will be shallower, if all disks are substantially
sub-maximal.  
This is considerably shallower
than the baryonic TF of \citet{mcgaugh00}, 
who advocate $L \propto V^{4 \pm 0.1}$.  We attribute the bulk of the 
difference to a difference in distance scales and stellar \ml normalization.

\acknowledgements

We thank Stacy McGaugh for asking us for improved 
stellar \ml estimates for spiral galaxies, which started us off on this 
project.  We thank all of the stellar population modelers 
who provided us with their models, and often spent 
a considerable effort helping 
us better understand the models.
We thank Roelof Bottema, Stephane Courteau, Stacy McGaugh and 
the anonymous referee for helpful comments.
E.F.~Bell was supported by NASA grant NAG5-8426 and 
NSF grant AST-9900789.
Support for R.S.~de Jong was provided by NASA through Hubble Fellowship
grant \#HF-01106.01-A from the Space Telescope Science Institute,
which is operated by the Association of Universities for Research in
Astronomy, Inc., under NASA contract NAS5-26555.

This research has made use of NASA's Astrophysics Data System Abstract
Service.  This research has made use of the NASA/IPAC Extragalactic
Database (NED) which is operated by the Jet Propulsion Laboratory,
California Institute of Technology, under contract with the National
Aeronautics and Space Administration.

%------------------------------------------------------------------------

\appendix

\section{Model tables}\label{sec:app}

%figure:acol
\begin{figure*}[tbh]
\epsfxsize=15cm \hspace{1cm}
\epsfbox{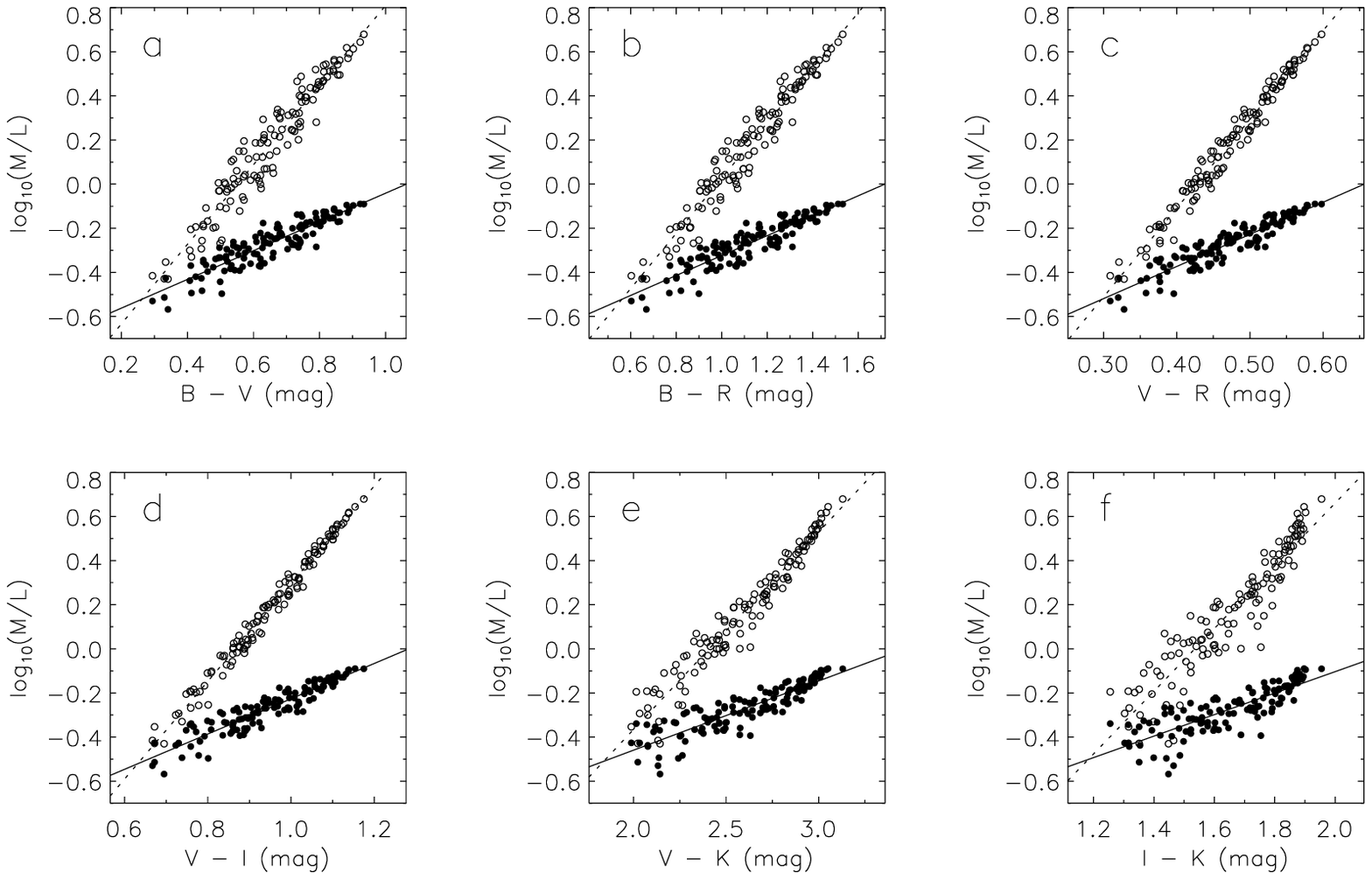}
\caption{\label{fig:acol} 
Trends in stellar M/L for the formation epoch model with bursts 
in $K$ (filled circles)
and $B$ band (open circles) with $B - V$ (a), $B - R$ (b), $V - R$ (c), 
$V - I$ (d), $V - K$ (e), and $I - K$ (f) color.  We also show the 
least-squares fit to the variations of stellar \ml with color 
for the $B$ (dotted line) and $K$ band stellar \ml (solid line).
 }
\end{figure*}

%figure:amod
\begin{figure*}[tbh]
\epsfxsize=15cm \hspace{1cm}
\epsfbox{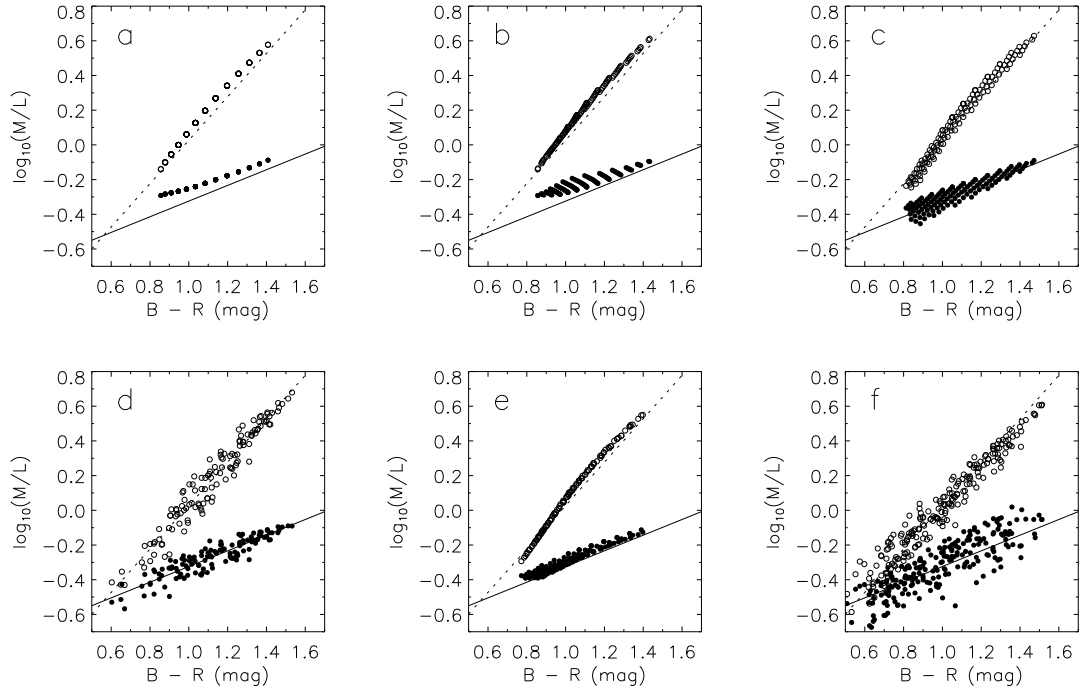}
\caption{\label{fig:amod} 
Trends in stellar M/L with $B - R$ color for six different galaxy 
evolution models in $K$ (filled circles)
and $B$ band (open circles) for the closed box model (a), outflow model (b), 
mass-dependent formation epoch model (c), mass-dependent formation 
epoch model with bursts (d), infall model (e), and \protect\citet{cole00} 
heirarchical model (f).  We also show the 
least-squares fit to the variations of stellar \ml with $B - R$ color of
the mass-dependent formation epoch with bursts model
for the $B$ (dotted line) and $K$ band (solid line).  
The \citet{cole00} model adopts a \citet{kennicutt83} IMF and
a 38\% brown dwarf fraction, which results in a similar zero point to the
scaled-down Salpeter IMF we adopt.  
}
\end{figure*}

In this appendix we present plots of the stellar \mls of the 
preferred model (the mass-dependent formation epoch model with bursts, 
with a scaled Salpeter IMF) against six colors (Fig.\ \ref{fig:acol}) and
plots of the stellar \ml against $B - R$ color for six different models
(Fig.\ \ref{fig:amod}).  We also present least-squares fits to the 
variation of stellar \ml of a variety of different galaxy evolution 
and SPS models with a wide range of colors (Tables 
\ref{tab:mlcol} and \ref{tab:mlsps} respectively).  Further discussion 
of these tables and figures are presented in the text.

\clearpage

\begin{table*}
\begin{footnotesize}
\begin{center}
\caption{Stellar M/L as a function of color for the scaled Salpeter IMF
	{\label{tab:mlcol}}} 
\begin{tabular}{lcccccccccccccc}
\tableline
\tableline
{Model} & {$a_B$} & {$b_B$} 
& {$a_V$} & {$b_V$} 
& {$a_R$} & {$b_R$} 
& {$a_I$} & {$b_I$} 
& {$a_J$} & {$b_J$} 
& {$a_H$} & {$b_H$} 
& {$a_K$} & {$b_K$} \\
\tableline
\multicolumn{15}{c}{$B - V$} \\
\tableline
Closed box & $-$1.019 & 1.937 & $-$0.759 & 1.537 & $-$0.681 & 1.346 & $-$0.631 & 1.170 & $-$0.540 & 0.767 & $-$0.553 & 0.632 & $-$0.554 & 0.540 \\
Infall & $-$1.113 & 2.065 & $-$0.853 & 1.665 & $-$0.772 & 1.468 & $-$0.723 & 1.290 & $-$0.658 & 0.907 & $-$0.679 & 0.777 & $-$0.692 & 0.699 \\ 
Outflow  & $-$1.026 & 1.954 & $-$0.766 & 1.554 & $-$0.685 & 1.357 & $-$0.634 & 1.179 & $-$0.527 & 0.741 & $-$0.536 & 0.600 & $-$0.534 & 0.500 \\ 
Dynamical time & $-$0.990 & 1.883 & $-$0.730 & 1.483 & $-$0.650 & 1.289 & $-$0.601 & 1.114 & $-$0.514 & 0.704 & $-$0.528 & 0.569 & $-$0.531 & 0.476 \\ 
Formation epoch & $-$1.110 & 2.018 & $-$0.850 & 1.618 & $-$0.770 & 1.425 & $-$0.724 & 1.257 & $-$0.659 & 0.878 & $-$0.683 & 0.757 & $-$0.694 & 0.676 \\ 
Form.\ epoch: bursts  & $-$0.994 & 1.804 & $-$0.734 & 1.404 & $-$0.660 & 1.222 & $-$0.627 & 1.075 & $-$0.621 & 0.794 & $-$0.663 & 0.704 & $-$0.692 & 0.652 \\
\protect\citet{cole00} & $-$0.888 & 1.758 & $-$0.628 & 1.358 & $-$0.565 & 1.132 & $-$0.525 & 0.981 & $-$0.550 & 0.801 & $-$0.618 & 0.718 & $-$0.654 & 0.696 \\ 
\tableline
\multicolumn{15}{c}{$B - R$} \\
\tableline
Closed box & $-$1.236 & 1.312 & $-$0.932 & 1.042 & $-$0.832 & 0.912 & $-$0.762 & 0.793 & $-$0.626 & 0.519 & $-$0.623 & 0.427 & $-$0.613 & 0.364 \\ 
Infall & $-$1.334 & 1.386 & $-$1.032 & 1.119 & $-$0.930 & 0.986 & $-$0.861 & 0.867 & $-$0.754 & 0.608 & $-$0.760 & 0.520 & $-$0.764 & 0.467 \\ 
Outflow  & $-$1.236 & 1.313 & $-$0.933 & 1.045 & $-$0.832 & 0.913 & $-$0.761 & 0.793 & $-$0.604 & 0.496 & $-$0.598 & 0.400 & $-$0.583 & 0.332 \\ 
Dynamical time & $-$1.195 & 1.270 & $-$0.892 & 1.001 & $-$0.791 & 0.870 & $-$0.723 & 0.752 & $-$0.590 & 0.474 & $-$0.589 & 0.382 & $-$0.581 & 0.319 \\ 
Formation epoch & $-$1.333 & 1.365 & $-$1.030 & 1.095 & $-$0.929 & 0.965 & $-$0.865 & 0.851 & $-$0.757 & 0.594 & $-$0.767 & 0.512 & $-$0.769 & 0.457 \\ 
Form.\ epoch: bursts  & $-$1.224 & 1.251 & $-$0.916 & 0.976 & $-$0.820 & 0.851 & $-$0.768 & 0.748 & $-$0.724 & 0.552 & $-$0.754 & 0.489 & $-$0.776 & 0.452 \\ 
\protect\citet{cole00} & $-$1.121 & 1.130 & $-$0.811 & 0.875 & $-$0.717 & 0.730 & $-$0.657 & 0.633 & $-$0.657 & 0.516 & $-$0.713 & 0.461 & $-$0.746 & 0.447 \\
\tableline
\multicolumn{15}{c}{$V - I$} \\
\tableline
Closed box & $-$1.771 & 2.104 & $-$1.359 & 1.674 & $-$1.207 & 1.466 & $-$1.087 & 1.274 & $-$0.835 & 0.830 & $-$0.791 & 0.679 & $-$0.755 & 0.578 \\
Infall & $-$1.882 & 2.191 & $-$1.478 & 1.772 & $-$1.323 & 1.563 & $-$1.206 & 1.372 & $-$0.988 & 0.954 & $-$0.955 & 0.810 & $-$0.935 & 0.723 \\
Outflow  & $-$1.743 & 2.072 & $-$1.341 & 1.653 & $-$1.188 & 1.445 & $-$1.069 & 1.253 & $-$0.786 & 0.772 & $-$0.737 & 0.615 & $-$0.692 & 0.503 \\ 
Dynamical time & $-$1.714 & 2.035 & $-$1.304 & 1.607 & $-$1.150 & 1.398 & $-$1.032 & 1.207 & $-$0.781 & 0.757 & $-$0.739 & 0.606 & $-$0.703 & 0.503 \\ 
Formation epoch & $-$1.931 & 2.234 & $-$1.513 & 1.797 & $-$1.356 & 1.584 & $-$1.241 & 1.397 & $-$1.017 & 0.972 & $-$0.989 & 0.835 & $-$0.965 & 0.744 \\ 
Form.\ epoch: bursts  & $-$1.919 & 2.214 & $-$1.476 & 1.747 & $-$1.314 & 1.528 & $-$1.204 & 1.347 & $-$1.040 & 0.987 & $-$1.030 & 0.870 & $-$1.027 & 0.800 \\ 
\protect\citet{cole00} & $-$1.674 & 1.865 & $-$1.249 & 1.456 & $-$1.088 & 1.220 & $-$0.977 & 1.056 & $-$0.901 & 0.843 & $-$0.924 & 0.745 & $-$0.943 & 0.714 \\ 
\tableline
\multicolumn{15}{c}{$V - J$} \\
\tableline
Closed box & $-$1.574 & 0.993 & $-$1.204 & 0.791 & $-$1.071 & 0.693 & $-$0.969 & 0.602 & $-$0.756 & 0.391 & $-$0.725 & 0.319 & $-$0.698 & 0.271 \\ 
Infall & $-$1.740 & 1.054 & $-$1.365 & 0.854 & $-$1.224 & 0.753 & $-$1.117 & 0.660 & $-$0.917 & 0.454 & $-$0.889 & 0.382 & $-$0.872 & 0.338 \\
Outflow  & $-$1.453 & 0.920 & $-$1.113 & 0.735 & $-$0.989 & 0.643 & $-$0.895 & 0.557 & $-$0.665 & 0.335 & $-$0.633 & 0.263 & $-$0.599 & 0.211 \\
Dynamical time & $-$1.524 & 0.952 & $-$1.156 & 0.753 & $-$1.022 & 0.655 & $-$0.921 & 0.566 & $-$0.708 & 0.353 & $-$0.679 & 0.281 & $-$0.651 & 0.232 \\
Formation epoch & $-$1.780 & 1.072 & $-$1.392 & 0.863 & $-$1.250 & 0.761 & $-$1.146 & 0.670 & $-$0.944 & 0.463 & $-$0.923 & 0.396 & $-$0.903 & 0.350 \\ 
Form.\ epoch: bursts  & $-$1.903 & 1.138 & $-$1.477 & 0.905 & $-$1.319 & 0.794 & $-$1.209 & 0.700 & $-$1.029 & 0.505 & $-$1.014 & 0.442 & $-$1.005 & 0.402 \\ 
\protect\citet{cole00} & $-$1.854 & 1.141 & $-$1.397 & 0.896 & $-$1.215 & 0.752 & $-$1.080 & 0.647 & $-$0.949 & 0.496 & $-$0.948 & 0.427 & $-$0.953 & 0.402 \\ 
\tableline
\multicolumn{15}{c}{$V - H$} \\
\tableline
Closed box & $-$1.782 & 0.840 & $-$1.371 & 0.669 & $-$1.217 & 0.586 & $-$1.096 & 0.509 & $-$0.838 & 0.330 & $-$0.791 & 0.269 & $-$0.753 & 0.228 \\ 
Infall & $-$1.962 & 0.889 & $-$1.546 & 0.721 & $-$1.384 & 0.636 & $-$1.257 & 0.557 & $-$1.011 & 0.382 & $-$0.966 & 0.321 & $-$0.939 & 0.284 \\
Outflow  & $-$1.641 & 0.776 & $-$1.264 & 0.621 & $-$1.121 & 0.543 & $-$1.009 & 0.470 & $-$0.731 & 0.282 & $-$0.684 & 0.221 & $-$0.639 & 0.176 \\ 
Dynamical time & $-$1.725 & 0.805 & $-$1.317 & 0.638 & $-$1.161 & 0.555 & $-$1.042 & 0.479 & $-$0.783 & 0.298 & $-$0.737 & 0.238 & $-$0.699 & 0.196 \\ 
Formation epoch & $-$2.027 & 0.916 & $-$1.592 & 0.737 & $-$1.425 & 0.650 & $-$1.300 & 0.572 & $-$1.050 & 0.395 & $-$1.012 & 0.337 & $-$0.981 & 0.298 \\ 
Form.\ epoch: bursts  & $-$2.181 & 0.978 & $-$1.700 & 0.779 & $-$1.515 & 0.684 & $-$1.383 & 0.603 & $-$1.151 & 0.434 & $-$1.120 & 0.379 & $-$1.100 & 0.345 \\ 
\protect\citet{cole00} & $-$2.142 & 0.961 & $-$1.627 & 0.756 & $-$1.410 & 0.635 & $-$1.246 & 0.546 & $-$1.070 & 0.415 & $-$1.047 & 0.356 & $-$1.043 & 0.333 \\ 
\tableline
\multicolumn{15}{c}{$V - K$} \\
\tableline
Closed box & $-$1.738 & 0.761 & $-$1.336 & 0.607 & $-$1.187 & 0.531 & $-$1.069 & 0.462 & $-$0.820 & 0.299 & $-$0.776 & 0.244 & $-$0.740 & 0.207 \\ 
Infall & $-$1.931 & 0.811 & $-$1.522 & 0.658 & $-$1.363 & 0.580 & $-$1.238 & 0.508 & $-$0.996 & 0.348 & $-$0.953 & 0.292 & $-$0.926 & 0.258 \\ 
Outflow  & $-$1.583 & 0.696 & $-$1.218 & 0.557 & $-$1.081 & 0.487 & $-$0.974 & 0.421 & $-$0.708 & 0.252 & $-$0.664 & 0.197 & $-$0.622 & 0.157 \\ 
Dynamical time & $-$1.682 & 0.728 & $-$1.283 & 0.577 & $-$1.132 & 0.502 & $-$1.016 & 0.433 & $-$0.766 & 0.270 & $-$0.723 & 0.215 & $-$0.687 & 0.177 \\ 
Formation epoch & $-$1.990 & 0.832 & $-$1.562 & 0.670 & $-$1.399 & 0.590 & $-$1.277 & 0.520 & $-$1.032 & 0.358 & $-$0.996 & 0.305 & $-$0.966 & 0.270 \\
Form.\ epoch: bursts  & $-$2.156 & 0.895 & $-$1.683 & 0.714 & $-$1.501 & 0.627 & $-$1.370 & 0.553 & $-$1.139 & 0.396 & $-$1.108 & 0.346 & $-$1.087 & 0.314 \\ 
\protect\citet{cole00} & $-$2.125 & 0.891 & $-$1.615 & 0.701 & $-$1.400 & 0.590 & $-$1.235 & 0.506 & $-$1.051 & 0.380 & $-$1.026 & 0.324 & $-$1.019 & 0.301 \\ 
\tableline \\
\end{tabular}
\end{center}
$\log_{10}({\rm M/L}) = a_{\lambda} + b_{\lambda} {\rm Color} $ \\ 
The \citet{cole00} model adopts a \citet{kennicutt83} IMF and
a 38\% brown dwarf fraction, which results in a similar zero point to the
scaled-down Salpeter IMF we adopt.  Note that the stellar \ml values can 
be estimated for any combination of the above colors by a simple linear
combination of the above fits.
\end{footnotesize}
\end{table*} 

\clearpage

\begin{table*}
\begin{footnotesize}
\begin{center}
\caption{Stellar M/L as a function of color for different SPS models
	{\label{tab:mlsps}}} 
\begin{tabular}{lcccccccccccccc}
\tableline
\tableline
{Model} & {$a_B$} & {$b_B$} 
& {$a_V$} & {$b_V$} 
& {$a_R$} & {$b_R$} 
& {$a_I$} & {$b_I$} 
& {$a_J$} & {$b_J$} 
& {$a_H$} & {$b_H$} 
& {$a_K$} & {$b_K$} \\
\tableline
\multicolumn{15}{c}{$B - V$ \hspace{2.0cm} $Z=0.008$} \\
\tableline
 Bruzual \& Charlot, Salpeter IMF           &$-$0.63& 1.54 &$-$0.37& 1.14 &$-$0.30& 0.97 &$-$0.27& 0.83 &$-$0.33& 0.68 &$-$0.39& 0.62 &$-$0.43& 0.60\\
 Bruzual \& Charlot, Scaled Salpeter IMF    &$-$0.78& 1.54 &$-$0.52& 1.14 &$-$0.46& 0.97 &$-$0.43& 0.83 &$-$0.48& 0.68 &$-$0.54& 0.62 &$-$0.59& 0.60\\
 Bruzual \& Charlot, Modified Salpeter IMF  &$-$0.91& 1.53 &$-$0.65& 1.13 &$-$0.58& 0.97 &$-$0.56& 0.84 &$-$0.61& 0.68 &$-$0.67& 0.61 &$-$0.71& 0.60\\
 Bruzual \& Charlot 96, Scalo IMF           &$-$0.85& 1.48 &$-$0.59& 1.08 &$-$0.52& 0.91 &$-$0.49& 0.79 &$-$0.56& 0.64 &$-$0.63& 0.57 &$-$0.65& 0.56\\
 Kodama \& Arimoto, Salpeter IMF            &$-$0.56& 1.46 &$-$0.30& 1.06 &$-$0.24& 0.91 &$-$0.22& 0.76 &$-$0.26& 0.61 &$-$0.36& 0.55 &$-$0.38& 0.53\\
 Schulz et al., Salpeter IMF                &$-$0.59& 1.55 &$-$0.33& 1.15 &$-$0.26& 0.98 &$-$0.29& 0.86 &$-$0.60& 0.76 &$-$0.51& 0.75 &$-$0.65& 0.77\\
 {\sc p\'egase}, Salpeter IMF               	    &$-$0.57& 1.51 &$-$0.31& 1.11 &$-$0.24& 0.95 &$-$0.19& 0.76 &$-$0.25& 0.65 &$-$0.33& 0.60 &$-$0.38& 0.59\\
 {\sc p\'egase}, $x=-1.85$ IMF       		    &$-$0.25& 1.40 & 0.01& 1.00 & 0.09& 0.82 & 0.11& 0.65 & 0.05& 0.55 &$-$0.03& 0.49 &$-$0.07& 0.48\\
 {\sc p\'egase}, $x=-0.85$ IMF       		    &$-$0.87& 1.75 &$-$0.61& 1.35 &$-$0.56& 1.20 &$-$0.42& 0.90 &$-$0.47& 0.79 &$-$0.53& 0.70 &$-$0.59& 0.71\\
\tableline
\multicolumn{15}{c}{$B - V$ \hspace{2.0cm} $Z=0.02$} \\
\tableline
 Bruzual \& Charlot, Salpeter IMF           &$-$0.51& 1.45 &$-$0.25& 1.05 &$-$0.19& 0.88 &$-$0.17& 0.76 &$-$0.28& 0.58 &$-$0.36& 0.53 &$-$0.42& 0.52\\
 Bruzual \& Charlot, Scaled Salpeter IMF    &$-$0.66& 1.45 &$-$0.40& 1.05 &$-$0.34& 0.88 &$-$0.33& 0.76 &$-$0.43& 0.58 &$-$0.51& 0.53 &$-$0.57& 0.52\\
 Bruzual \& Charlot, Modified Salpeter IMF  &$-$0.79& 1.43 &$-$0.53& 1.03 &$-$0.47& 0.87 &$-$0.46& 0.75 &$-$0.55& 0.58 &$-$0.63& 0.53 &$-$0.69& 0.51\\
 Bruzual \& Charlot 96, Scalo IMF           &$-$0.74& 1.40 &$-$0.48& 1.00 &$-$0.42& 0.84 &$-$0.40& 0.73 &$-$0.49& 0.51 &$-$0.58& 0.45 &$-$0.61& 0.43\\
 Kodama \& Arimoto, Salpeter IMF            &$-$0.44& 1.40 &$-$0.18& 1.00 &$-$0.12& 0.84 &$-$0.12& 0.70 &$-$0.19& 0.53 &$-$0.30& 0.48 &$-$0.33& 0.45\\
 Schulz et al., Salpeter IMF                &$-$0.49& 1.46 &$-$0.23& 1.06 &$-$0.16& 0.89 &$-$0.20& 0.77 &$-$0.59& 0.61 &$-$0.48& 0.61 &$-$0.64& 0.62\\
 {\sc p\'egase}, Salpeter IMF               &$-$0.47& 1.45 &$-$0.21& 1.05 &$-$0.15& 0.89 &$-$0.12& 0.71 &$-$0.23& 0.59 &$-$0.34& 0.55 &$-$0.39& 0.54\\
 {\sc p\'egase}, $x=-1.85$ IMF       &$-$0.15& 1.36 & 0.11& 0.96 & 0.18& 0.79 & 0.19& 0.62 & 0.09& 0.49 &$-$0.02& 0.44 &$-$0.07& 0.44\\
 {\sc p\'egase}, $x=-0.85$ IMF       &$-$0.77& 1.65 &$-$0.51& 1.25 &$-$0.47& 1.11 &$-$0.35& 0.85 &$-$0.46& 0.72 &$-$0.55& 0.65 &$-$0.62& 0.65\\
\tableline
\multicolumn{15}{c}{$B - R$ \hspace{2.0cm} $Z=0.008$} \\
\tableline
 Bruzual \& Charlot, Salpeter IMF           &$-$0.84& 1.08 &$-$0.52& 0.80 &$-$0.43& 0.68 &$-$0.39& 0.59 &$-$0.42& 0.48 &$-$0.47& 0.43 &$-$0.51& 0.42\\
 Bruzual \& Charlot, Scaled Salpeter IMF    &$-$0.99& 1.08 &$-$0.68& 0.80 &$-$0.59& 0.68 &$-$0.54& 0.59 &$-$0.57& 0.48 &$-$0.63& 0.43 &$-$0.67& 0.42\\
 Bruzual \& Charlot, Modified Salpeter IMF  &$-$1.12& 1.08 &$-$0.81& 0.80 &$-$0.72& 0.68 &$-$0.68& 0.59 &$-$0.70& 0.48 &$-$0.75& 0.44 &$-$0.80& 0.42\\
 Bruzual \& Charlot 96, Scalo IMF           &$-$1.05& 1.04 &$-$0.74& 0.76 &$-$0.65& 0.64 &$-$0.60& 0.56 &$-$0.64& 0.45 &$-$0.70& 0.40 &$-$0.73& 0.39\\
 Kodama \& Arimoto, Salpeter IMF            &$-$0.77& 1.05 &$-$0.45& 0.76 &$-$0.36& 0.65 &$-$0.33& 0.55 &$-$0.35& 0.44 &$-$0.44& 0.40 &$-$0.45& 0.38\\
 Schulz et al., Salpeter IMF                &$-$0.79& 1.08 &$-$0.48& 0.80 &$-$0.39& 0.68 &$-$0.40& 0.60 &$-$0.70& 0.53 &$-$0.61& 0.52 &$-$0.75& 0.54\\
 {\sc p\'egase}, Salpeter IMF               &$-$0.78& 1.08 &$-$0.47& 0.79 &$-$0.38& 0.68 &$-$0.30& 0.54 &$-$0.35& 0.46 &$-$0.42& 0.42 &$-$0.47& 0.42\\
 {\sc p\'egase}, $x=-1.85$ IMF       &$-$0.42& 0.97 &$-$0.11& 0.70 &$-$0.02& 0.57 & 0.02& 0.46 &$-$0.02& 0.38 &$-$0.09& 0.34 &$-$0.13& 0.33\\
 {\sc p\'egase}, $x=-0.85$ IMF       &$-$1.18& 1.29 &$-$0.85& 0.99 &$-$0.77& 0.89 &$-$0.58& 0.66 &$-$0.61& 0.58 &$-$0.65& 0.52 &$-$0.72& 0.52\\
\tableline
\multicolumn{15}{c}{$B - R$ \hspace{2.0cm} $Z=0.02$} \\
\tableline
 Bruzual \& Charlot, Salpeter IMF           &$-$0.73& 1.03 &$-$0.41& 0.74 &$-$0.32& 0.63 &$-$0.29& 0.54 &$-$0.36& 0.41 &$-$0.44& 0.38 &$-$0.49& 0.37\\
 Bruzual \& Charlot, Scaled Salpeter IMF    &$-$0.88& 1.03 &$-$0.56& 0.74 &$-$0.48& 0.63 &$-$0.44& 0.54 &$-$0.52& 0.41 &$-$0.59& 0.38 &$-$0.65& 0.37\\
 Bruzual \& Charlot, Modified Salpeter IMF  &$-$1.01& 1.02 &$-$0.69& 0.74 &$-$0.60& 0.62 &$-$0.57& 0.54 &$-$0.64& 0.41 &$-$0.71& 0.38 &$-$0.77& 0.37\\
 Bruzual \& Charlot 96, Scalo IMF           &$-$0.95& 1.00 &$-$0.63& 0.71 &$-$0.54& 0.60 &$-$0.50& 0.52 &$-$0.56& 0.37 &$-$0.65& 0.32 &$-$0.67& 0.31\\
 Kodama \& Arimoto, Salpeter IMF            &$-$0.65& 1.00 &$-$0.33& 0.72 &$-$0.25& 0.60 &$-$0.22& 0.50 &$-$0.27& 0.38 &$-$0.38& 0.35 &$-$0.40& 0.32\\
 Schulz et al., Salpeter IMF                &$-$0.69& 1.02 &$-$0.38& 0.74 &$-$0.29& 0.62 &$-$0.31& 0.54 &$-$0.67& 0.43 &$-$0.57& 0.43 &$-$0.73& 0.44\\
 {\sc p\'egase}, Salpeter IMF               &$-$0.70& 1.04 &$-$0.38& 0.75 &$-$0.29& 0.64 &$-$0.23& 0.51 &$-$0.32& 0.43 &$-$0.42& 0.39 &$-$0.48& 0.39\\
 {\sc p\'egase}, $x=-1.85$ IMF       &$-$0.33& 0.95 &$-$0.02& 0.67 & 0.07& 0.55 & 0.10& 0.44 & 0.02& 0.34 &$-$0.08& 0.31 &$-$0.13& 0.31\\
 {\sc p\'egase}, $x=-0.85$ IMF       &$-$1.08& 1.22 &$-$0.74& 0.92 &$-$0.67& 0.82 &$-$0.51& 0.62 &$-$0.59& 0.53 &$-$0.67& 0.48 &$-$0.74& 0.48\\
\tableline
\multicolumn{15}{c}{$V - I$ \hspace{2.0cm} $Z=0.008$} \\
\tableline
 Bruzual \& Charlot, Salpeter IMF           &$-$1.53& 2.01 &$-$1.04& 1.49 &$-$0.87& 1.27 &$-$0.77& 1.09 &$-$0.73& 0.88 &$-$0.75& 0.80 &$-$0.78& 0.78\\
 Bruzual \& Charlot, Scaled Salpeter IMF    &$-$1.69& 2.01 &$-$1.19& 1.49 &$-$1.03& 1.27 &$-$0.92& 1.09 &$-$0.88& 0.88 &$-$0.91& 0.80 &$-$0.94& 0.78\\
 Bruzual \& Charlot, Modified Salpeter IMF  &$-$1.85& 2.07 &$-$1.34& 1.53 &$-$1.18& 1.31 &$-$1.07& 1.13 &$-$1.02& 0.92 &$-$1.05& 0.83 &$-$1.08& 0.81\\
 Bruzual \& Charlot 96, Scalo IMF           &$-$1.76& 2.10 &$-$1.25& 1.53 &$-$1.08& 1.29 &$-$0.98& 1.13 &$-$0.95& 0.91 &$-$0.98& 0.81 &$-$1.00& 0.79\\
 Kodama \& Arimoto, Salpeter IMF            &$-$1.49& 1.95 &$-$0.98& 1.42 &$-$0.81& 1.21 &$-$0.71& 1.02 &$-$0.65& 0.81 &$-$0.71& 0.74 &$-$0.71& 0.70\\
 Schulz et al., Salpeter IMF                &$-$1.82& 2.15 &$-$1.25& 1.60 &$-$1.04& 1.36 &$-$0.98& 1.20 &$-$1.20& 1.05 &$-$1.10& 1.03 &$-$1.26& 1.07\\
 {\sc p\'egase}, Salpeter IMF               &$-$1.25& 1.70 &$-$0.81& 1.25 &$-$0.67& 1.07 &$-$0.54& 0.85 &$-$0.55& 0.74 &$-$0.60& 0.67 &$-$0.65& 0.67\\
 {\sc p\'egase}, $x=-1.85$ IMF       &$-$0.98& 1.62 &$-$0.51& 1.16 &$-$0.34& 0.96 &$-$0.24& 0.76 &$-$0.23& 0.63 &$-$0.29& 0.57 &$-$0.32& 0.56\\
 {\sc p\'egase}, $x=-0.85$ IMF       &$-$1.18& 1.57 &$-$0.85& 1.22 &$-$0.78& 1.09 &$-$0.58& 0.82 &$-$0.61& 0.71 &$-$0.66& 0.64 &$-$0.72& 0.64\\
\tableline
\multicolumn{15}{c}{$V - I$ \hspace{2.0cm} $Z=0.02$} \\
\tableline
 Bruzual \& Charlot, Salpeter IMF           &$-$1.50& 1.99 &$-$0.97& 1.44 &$-$0.80& 1.21 &$-$0.70& 1.04 &$-$0.68& 0.80 &$-$0.72& 0.73 &$-$0.77& 0.71\\
 Bruzual \& Charlot, Scaled Salpeter IMF    &$-$1.65& 1.99 &$-$1.12& 1.44 &$-$0.95& 1.21 &$-$0.85& 1.04 &$-$0.83& 0.80 &$-$0.88& 0.73 &$-$0.93& 0.71\\
 Bruzual \& Charlot, Modified Salpeter IMF  &$-$1.81& 2.04 &$-$1.27& 1.47 &$-$1.09& 1.24 &$-$0.99& 1.07 &$-$0.97& 0.83 &$-$1.01& 0.75 &$-$1.06& 0.73\\
 Bruzual \& Charlot 96, Scalo IMF           &$-$1.71& 2.06 &$-$1.17& 1.47 &$-$1.00& 1.24 &$-$0.90& 1.07 &$-$0.84& 0.75 &$-$0.89& 0.67 &$-$0.91& 0.63\\
 Kodama \& Arimoto, Salpeter IMF            &$-$1.42& 1.87 &$-$0.88& 1.34 &$-$0.71& 1.13 &$-$0.61& 0.94 &$-$0.57& 0.71 &$-$0.64& 0.65 &$-$0.65& 0.61\\
 Schulz et al., Salpeter IMF                &$-$1.73& 2.02 &$-$1.13& 1.47 &$-$0.92& 1.23 &$-$0.86& 1.07 &$-$1.11& 0.84 &$-$1.01& 0.85 &$-$1.18& 0.86\\
 {\sc p\'egase}, Salpeter IMF               &$-$1.25& 1.72 &$-$0.78& 1.25 &$-$0.63& 1.06 &$-$0.50& 0.85 &$-$0.55& 0.70 &$-$0.63& 0.65 &$-$0.68& 0.64\\
 {\sc p\'egase}, $x=-1.85$ IMF       &$-$0.93& 1.61 &$-$0.44& 1.14 &$-$0.28& 0.93 &$-$0.17& 0.74 &$-$0.20& 0.58 &$-$0.27& 0.53 &$-$0.32& 0.52\\
 {\sc p\'egase}, $x=-0.85$ IMF       &$-$1.24& 1.63 &$-$0.87& 1.24 &$-$0.79& 1.10 &$-$0.60& 0.84 &$-$0.66& 0.71 &$-$0.74& 0.65 &$-$0.81& 0.65\\
\end{tabular}
\end{center}
\end{footnotesize}
\end{table*} 

\begin{table*}
\tablenum{A4}
\begin{footnotesize}
\begin{center}
\caption{Continued} 
\begin{tabular}{lcccccccccccccc}
\tableline
\tableline
{Model} & {$a_B$} & {$b_B$} 
& {$a_V$} & {$b_V$} 
& {$a_R$} & {$b_R$} 
& {$a_I$} & {$b_I$} 
& {$a_J$} & {$b_J$} 
& {$a_H$} & {$b_H$} 
& {$a_K$} & {$b_K$} \\
\tableline
\multicolumn{15}{c}{$V - J$ \hspace{2.0cm} $Z=0.008$} \\
\tableline
 Bruzual \& Charlot, Salpeter IMF           &$-$1.98& 1.32 &$-$1.37& 0.98 &$-$1.15& 0.84 &$-$1.01& 0.72 &$-$0.92& 0.58 &$-$0.93& 0.53 &$-$0.96& 0.51\\
 Bruzual \& Charlot, Scaled Salpeter IMF    &$-$2.13& 1.32 &$-$1.52& 0.98 &$-$1.31& 0.84 &$-$1.16& 0.72 &$-$1.08& 0.58 &$-$1.08& 0.53 &$-$1.11& 0.51\\
 Bruzual \& Charlot, Modified Salpeter IMF  &$-$2.28& 1.35 &$-$1.66& 1.00 &$-$1.45& 0.85 &$-$1.31& 0.74 &$-$1.22& 0.60 &$-$1.22& 0.54 &$-$1.25& 0.53\\
 Bruzual \& Charlot 96, Scalo IMF           &$-$2.24& 1.35 &$-$1.60& 0.98 &$-$1.38& 0.83 &$-$1.24& 0.73 &$-$1.15& 0.58 &$-$1.16& 0.52 &$-$1.18& 0.51\\
 Kodama \& Arimoto, Salpeter IMF            &$-$1.87& 1.29 &$-$1.26& 0.94 &$-$1.05& 0.80 &$-$0.91& 0.67 &$-$0.81& 0.54 &$-$0.85& 0.49 &$-$0.85& 0.46\\
 Schulz et al., Salpeter IMF                &$-$3.40& 1.58 &$-$2.42& 1.17 &$-$2.04& 1.00 &$-$1.86& 0.88 &$-$1.97& 0.77 &$-$1.86& 0.76 &$-$2.05& 0.79\\
 {\sc p\'egase}, Salpeter IMF               &$-$1.86& 1.31 &$-$1.26& 0.97 &$-$1.06& 0.83 &$-$0.85& 0.66 &$-$0.82& 0.57 &$-$0.85& 0.52 &$-$0.89& 0.51\\
 {\sc p\'egase}, $x=-1.85$ IMF       &$-$1.52& 1.23 &$-$0.89& 0.88 &$-$0.66& 0.73 &$-$0.49& 0.58 &$-$0.45& 0.48 &$-$0.48& 0.43 &$-$0.51& 0.42\\
 {\sc p\'egase}, $x=-0.85$ IMF       &$-$1.82& 1.24 &$-$1.35& 0.96 &$-$1.23& 0.86 &$-$0.92& 0.65 &$-$0.90& 0.56 &$-$0.92& 0.51 &$-$0.99& 0.51\\
\tableline
\multicolumn{15}{c}{$V - J$ \hspace{2.0cm} $Z=0.02$} \\
\tableline
 Bruzual \& Charlot, Salpeter IMF           &$-$1.99& 1.24 &$-$1.32& 0.90 &$-$1.09& 0.76 &$-$0.95& 0.65 &$-$0.87& 0.50 &$-$0.90& 0.46 &$-$0.95& 0.45\\
 Bruzual \& Charlot, Scaled Salpeter IMF    &$-$2.14& 1.24 &$-$1.48& 0.90 &$-$1.25& 0.76 &$-$1.11& 0.65 &$-$1.03& 0.50 &$-$1.06& 0.46 &$-$1.10& 0.45\\
 Bruzual \& Charlot, Modified Salpeter IMF  &$-$2.29& 1.27 &$-$1.62& 0.92 &$-$1.39& 0.77 &$-$1.25& 0.67 &$-$1.17& 0.52 &$-$1.19& 0.47 &$-$1.23& 0.46\\
 Bruzual \& Charlot 96, Scalo IMF           &$-$2.04& 1.15 &$-$1.41& 0.82 &$-$1.20& 0.69 &$-$1.08& 0.60 &$-$0.97& 0.42 &$-$1.00& 0.37 &$-$1.01& 0.35\\
 Kodama \& Arimoto, Salpeter IMF            &$-$1.82& 1.20 &$-$1.17& 0.86 &$-$0.95& 0.72 &$-$0.81& 0.60 &$-$0.72& 0.46 &$-$0.78& 0.41 &$-$0.78& 0.39\\
 Schulz et al., Salpeter IMF                &$-$3.11& 1.30 &$-$2.13& 0.94 &$-$1.76& 0.79 &$-$1.59& 0.68 &$-$1.68& 0.54 &$-$1.59& 0.55 &$-$1.76& 0.55\\
 {\sc p\'egase}, Salpeter IMF               &$-$1.94& 1.26 &$-$1.28& 0.92 &$-$1.06& 0.78 &$-$0.85& 0.62 &$-$0.83& 0.52 &$-$0.89& 0.48 &$-$0.94& 0.47\\
 {\sc p\'egase}, $x=-1.85$ IMF       &$-$1.52& 1.16 &$-$0.86& 0.82 &$-$0.62& 0.67 &$-$0.44& 0.53 &$-$0.41& 0.42 &$-$0.46& 0.38 &$-$0.51& 0.37\\
 {\sc p\'egase}, $x=-0.85$ IMF       &$-$1.98& 1.23 &$-$1.44& 0.94 &$-$1.29& 0.83 &$-$0.98& 0.64 &$-$0.99& 0.54 &$-$1.04& 0.49 &$-$1.11& 0.49\\
\tableline
\multicolumn{15}{c}{$V - H$ \hspace{2.0cm} $Z=0.008$} \\
\tableline
 Bruzual \& Charlot, Salpeter IMF           &$-$2.39& 1.17 &$-$1.67& 0.87 &$-$1.42& 0.74 &$-$1.23& 0.64 &$-$1.10& 0.51 &$-$1.09& 0.47 &$-$1.12& 0.45\\
 Bruzual \& Charlot, Scaled Salpeter IMF    &$-$2.54& 1.17 &$-$1.83& 0.87 &$-$1.57& 0.74 &$-$1.39& 0.64 &$-$1.26& 0.51 &$-$1.25& 0.47 &$-$1.27& 0.45\\
 Bruzual \& Charlot, Modified Salpeter IMF  &$-$2.68& 1.18 &$-$1.96& 0.88 &$-$1.70& 0.75 &$-$1.53& 0.65 &$-$1.39& 0.52 &$-$1.38& 0.48 &$-$1.40& 0.46\\
 Bruzual \& Charlot 96, Scalo IMF           &$-$2.64& 1.16 &$-$1.90& 0.85 &$-$1.63& 0.72 &$-$1.46& 0.63 &$-$1.33& 0.50 &$-$1.32& 0.45 &$-$1.33& 0.44\\
 Kodama \& Arimoto, Salpeter IMF            &$-$2.38& 1.14 &$-$1.63& 0.83 &$-$1.36& 0.71 &$-$1.17& 0.60 &$-$1.02& 0.48 &$-$1.05& 0.43 &$-$1.03& 0.41\\
 Schulz et al., Salpeter IMF                &$-$3.48& 1.53 &$-$2.48& 1.13 &$-$2.08& 0.96 &$-$1.90& 0.85 &$-$2.01& 0.75 &$-$1.90& 0.73 &$-$2.08& 0.76\\
 {\sc p\'egase}, Salpeter IMF               &$-$2.34& 1.17 &$-$1.61& 0.86 &$-$1.36& 0.74 &$-$1.09& 0.59 &$-$1.02& 0.51 &$-$1.03& 0.46 &$-$1.08& 0.46\\
 {\sc p\'egase}, $x=-1.85$ IMF       &$-$1.96& 1.10 &$-$1.21& 0.79 &$-$0.93& 0.65 &$-$0.70& 0.52 &$-$0.62& 0.43 &$-$0.63& 0.39 &$-$0.66& 0.38\\
 {\sc p\'egase}, $x=-0.85$ IMF       &$-$2.23& 1.08 &$-$1.67& 0.84 &$-$1.51& 0.75 &$-$1.13& 0.57 &$-$1.09& 0.49 &$-$1.09& 0.44 &$-$1.15& 0.45\\
\tableline
\multicolumn{15}{c}{$V - H$ \hspace{2.0cm} $Z=0.02$} \\
\tableline
 Bruzual \& Charlot, Salpeter IMF           &$-$2.44& 1.12 &$-$1.65& 0.81 &$-$1.37& 0.69 &$-$1.19& 0.59 &$-$1.06& 0.45 &$-$1.07& 0.41 &$-$1.11& 0.40\\
 Bruzual \& Charlot, scaled Salpeter IMF    &$-$2.59& 1.12 &$-$1.80& 0.81 &$-$1.53& 0.69 &$-$1.34& 0.59 &$-$1.21& 0.45 &$-$1.22& 0.41 &$-$1.26& 0.40\\
 Bruzual \& Charlot, Modified Salpeter IMF  &$-$2.73& 1.14 &$-$1.93& 0.82 &$-$1.66& 0.69 &$-$1.48& 0.60 &$-$1.35& 0.46 &$-$1.35& 0.42 &$-$1.39& 0.41\\
 Bruzual \& Charlot 96, Scalo IMF           &$-$2.48& 1.03 &$-$1.73& 0.73 &$-$1.47& 0.62 &$-$1.30& 0.54 &$-$1.12& 0.38 &$-$1.15& 0.33 &$-$1.15& 0.32\\
 Kodama \& Arimoto, Salpeter IMF            &$-$2.34& 1.08 &$-$1.54& 0.77 &$-$1.26& 0.65 &$-$1.07& 0.54 &$-$0.92& 0.41 &$-$0.96& 0.37 &$-$0.95& 0.35\\
 Schulz et al., Salpeter IMF                &$-$3.22& 1.31 &$-$2.21& 0.95 &$-$1.83& 0.80 &$-$1.65& 0.69 &$-$1.73& 0.55 &$-$1.63& 0.55 &$-$1.81& 0.56\\
 {\sc p\'egase}, Salpeter IMF               &$-$2.48& 1.14 &$-$1.67& 0.83 &$-$1.40& 0.71 &$-$1.11& 0.56 &$-$1.06& 0.47 &$-$1.09& 0.43 &$-$1.14& 0.43\\
 {\sc p\'egase}, $x=-1.85$ IMF       &$-$2.02& 1.06 &$-$1.21& 0.75 &$-$0.91& 0.61 &$-$0.67& 0.48 &$-$0.59& 0.38 &$-$0.63& 0.35 &$-$0.67& 0.34\\
 {\sc p\'egase}, $x=-0.85$ IMF       &$-$2.47& 1.10 &$-$1.82& 0.84 &$-$1.63& 0.74 &$-$1.24& 0.57 &$-$1.21& 0.48 &$-$1.24& 0.44 &$-$1.31& 0.44\\
\tableline
\multicolumn{15}{c}{$V - K$ \hspace{2.0cm} $Z=0.008$} \\
\tableline
 Bruzual \& Charlot, Salpeter IMF           &$-$2.50& 1.13 &$-$1.75& 0.84 &$-$1.48& 0.71 &$-$1.29& 0.61 &$-$1.15& 0.50 &$-$1.14& 0.45 &$-$1.16& 0.44\\
 Bruzual \& Charlot, Scaled Salpeter IMF    &$-$2.65& 1.13 &$-$1.91& 0.84 &$-$1.64& 0.71 &$-$1.45& 0.61 &$-$1.30& 0.50 &$-$1.29& 0.45 &$-$1.31& 0.44\\
 Bruzual \& Charlot, Modified Salpeter IMF  &$-$2.79& 1.14 &$-$2.04& 0.85 &$-$1.77& 0.72 &$-$1.59& 0.62 &$-$1.44& 0.51 &$-$1.42& 0.46 &$-$1.45& 0.45\\
 Bruzual \& Charlot 96, Scalo IMF           &$-$2.72& 1.14 &$-$1.95& 0.83 &$-$1.68& 0.70 &$-$1.50& 0.61 &$-$1.36& 0.49 &$-$1.35& 0.44 &$-$1.36& 0.43\\
 Kodama \& Arimoto, Salpeter IMF            &$-$2.38& 1.09 &$-$1.63& 0.79 &$-$1.36& 0.67 &$-$1.17& 0.57 &$-$1.02& 0.45 &$-$1.05& 0.41 &$-$1.03& 0.39\\
 Schulz et al., Salpeter IMF                &$-$4.31& 1.63 &$-$3.09& 1.21 &$-$2.61& 1.03 &$-$2.36& 0.91 &$-$2.42& 0.80 &$-$2.30& 0.79 &$-$2.50& 0.81\\
 {\sc p\'egase}, Salpeter IMF               &$-$2.51& 1.16 &$-$1.74& 0.85 &$-$1.47& 0.73 &$-$1.17& 0.58 &$-$1.10& 0.50 &$-$1.10& 0.46 &$-$1.15& 0.45\\
 {\sc p\'egase}, $x=-1.85$ IMF       &$-$2.09& 1.08 &$-$1.30& 0.77 &$-$1.00& 0.64 &$-$0.76& 0.50 &$-$0.67& 0.42 &$-$0.68& 0.38 &$-$0.71& 0.37\\
 {\sc p\'egase}, $x=-0.85$ IMF       &$-$2.45& 1.10 &$-$1.84& 0.85 &$-$1.66& 0.76 &$-$1.25& 0.57 &$-$1.19& 0.50 &$-$1.18& 0.45 &$-$1.25& 0.45\\
\tableline
\multicolumn{15}{c}{$V - K$ \hspace{2.0cm} $Z=0.02$} \\
\tableline
 Bruzual \& Charlot, Salpeter IMF           &$-$2.59& 1.09 &$-$1.76& 0.79 &$-$1.46& 0.67 &$-$1.27& 0.57 &$-$1.12& 0.44 &$-$1.12& 0.40 &$-$1.16& 0.39\\
 Bruzual \& Charlot, Scaled Salpeter IMF    &$-$2.74& 1.09 &$-$1.91& 0.79 &$-$1.62& 0.67 &$-$1.42& 0.57 &$-$1.27& 0.44 &$-$1.28& 0.40 &$-$1.32& 0.39\\
 Bruzual \& Charlot, Modified Salpeter IMF  &$-$2.88& 1.10 &$-$2.04& 0.80 &$-$1.75& 0.67 &$-$1.56& 0.58 &$-$1.41& 0.45 &$-$1.41& 0.41 &$-$1.45& 0.40\\
 Bruzual \& Charlot 96, Scalo IMF           &$-$2.53& 0.98 &$-$1.76& 0.70 &$-$1.49& 0.59 &$-$1.33& 0.51 &$-$1.14& 0.36 &$-$1.16& 0.32 &$-$1.16& 0.30\\
 Kodama \& Arimoto, Salpeter IMF            &$-$2.35& 1.02 &$-$1.55& 0.73 &$-$1.27& 0.62 &$-$1.08& 0.51 &$-$0.92& 0.39 &$-$0.97& 0.35 &$-$0.96& 0.33\\
 Schulz et al., Salpeter IMF                &$-$3.87& 1.34 &$-$2.68& 0.97 &$-$2.23& 0.81 &$-$1.99& 0.71 &$-$2.00& 0.56 &$-$1.91& 0.56 &$-$2.09& 0.57\\
 {\sc p\'egase}, Salpeter IMF               &$-$2.67& 1.13 &$-$1.81& 0.82 &$-$1.51& 0.70 &$-$1.21& 0.56 &$-$1.13& 0.46 &$-$1.17& 0.43 &$-$1.22& 0.42\\
 {\sc p\'egase}, $x=-1.85$ IMF       &$-$2.17& 1.04 &$-$1.31& 0.73 &$-$0.99& 0.60 &$-$0.73& 0.48 &$-$0.64& 0.38 &$-$0.68& 0.34 &$-$0.72& 0.33\\
 {\sc p\'egase}, $x=-0.85$ IMF       &$-$2.71& 1.10 &$-$1.99& 0.84 &$-$1.78& 0.75 &$-$1.36& 0.57 &$-$1.31& 0.48 &$-$1.33& 0.44 &$-$1.40& 0.44\\
\tableline \\
\end{tabular}
\end{center}
$\log_{10}({\rm M/L}) = a_{\lambda} + b_{\lambda} {\rm Color} $ \\ 
Note that the stellar \ml values can 
be estimated for any combination of the above colors by a simple linear
combination of the above fits.
\end{footnotesize}
\end{table*} 

\end{document}